\documentclass[aps,
superscriptaddress,
nofootinbib,
longbibliography,
amsmath,
amssymb,
10pt
]{revtex4-2} 
\usepackage{mathtools}
\usepackage{amsfonts}
\usepackage{mathrsfs}
\usepackage{bbm}
\usepackage{slashed}
\usepackage{xspace}
\usepackage{amssymb}
\usepackage{cuted}
\usepackage{siunitx}
\usepackage{xspace}
\usepackage[normalem]{ulem}
\usepackage{amsmath, amssymb}
\usepackage{mathtools}
\usepackage{amsmath}
\usepackage[utf8]{inputenc}
\usepackage{graphicx}
\usepackage{cancel}
\usepackage{bbm}
\usepackage{bbold}
\usepackage{color}
\usepackage{comment}
\usepackage{lineno}
\usepackage{dcolumn}   
\usepackage{bm}       
\usepackage{verbatim}
\usepackage[T1]{fontenc}
\usepackage{lipsum}
\usepackage{subcaption}
\usepackage{caption}
\usepackage{float}
\usepackage[newcommands]{ragged2e}
\usepackage{graphicx,caption}
\usepackage{tensor}
\usepackage{mathtools}
\let\mathcal\relax
\DeclareMathAlphabet{\mathcal}{OMS}{cmsy}{m}{n}

\usepackage[dvipsnames]{xcolor}
\usepackage[colorlinks = true,
            linkcolor  = RoyalBlue,
            urlcolor   = RoyalBlue,
            citecolor  = RoyalBlue]{hyperref}

\newcommand{\fluidum}{Fluid\textit{u}M\xspace}
\newcommand{\trento}{T\raisebox{-.5ex}{R}ENTo }

\newcommand{\PbPb}         {\mbox{Pb--Pb}\xspace}

\newcommand{\pt}           {\ensuremath{p_{\rm T}}\xspace}

\newcommand{\fivenn}       {$\sqrt{s_{\mathrm{NN}}}=5.02$~TeV\xspace}

\begin{document}
\title{Out-of-equilibrium contributions to charm hadrons in a fluid-dynamic approach}

\author{Rossana Facen}
\email{r.facen@gsi.de}
\affiliation{Physikalisches Institut, Universität Heidelberg, 69120 Heidelberg, Germany}

\author{Federica Capellino}
\email{f.capellino@gsi.de}
\affiliation{GSI Helmoltzzentrum für Schwerionenforschung,
Planckstrasse 1,  64291 Darmstadt, Germany}

\author{Eduardo Grossi}
\email{eduardo.grossi@unifi.it }
\affiliation{Dipartimento di Fisica, Universit\`a di Firenze and INFN Sezione di Firenze, via G. Sansone 1,
50019 Sesto Fiorentino, Italy}

\author{Andrea Dubla}
\email{a.dubla@gsi.de}
\affiliation{GSI Helmoltzzentrum für Schwerionenforschung,
Planckstrasse 1,  64291 Darmstadt, Germany}

\author{Silvia Masciocchi}
\email{s.masciocchi@gsi.de}
\affiliation{Physikalisches Institut, Universität Heidelberg, 69120 Heidelberg, Germany}
\affiliation{GSI Helmoltzzentrum für Schwerionenforschung,
Planckstrasse 1,  64291 Darmstadt, Germany}

\date{\today}

\begin{abstract}
Building on previous studies that demonstrated the applicability of a fluid-dynamic description of charm quarks in the quark-gluon plasma \cite{Capellino:2022nvf,Capellino:2023cxe}, the present work extends the framework by computing the out-of-equilibrium contributions to the distribution function of charm hadrons. The analysis includes corrections arising from the initial out-of-equilibrium distribution of charm quarks after a free-streaming phase, as well as from the freeze-out surface within fluid dynamics. These results enable the exact computation of integrated yields and transverse momentum distributions of charm hadrons for different values of the spatial diffusion coefficient, thereby providing the basis for a systematic determination of the charm transport coefficients. A preliminary comparison of our model with the available experimental data shows compatibility within uncertainties. In addition, the limits of applicability of the approach are identified by determining the transverse-momentum region in which charm hadrons are described by a well-defined, positive distribution function.
\end{abstract}

\maketitle

\section{Introduction}
\label{sec:fluid_dynamic_approach}
Charm quarks represent unique probes to study the properties of the hot and dense QCD medium produced in high-energy heavy-ion collisions, the quark-gluon plasma (QGP). 
Since their mass is significantly larger than the typical temperature of the QGP reached in nuclear collision experiments, charm quarks are produced almost exclusively in the initial stage via hard partonic scatterings, and their thermal production in the fireball is exponentially suppressed~\cite{Braun-Munzinger:2007fth}. Moreover, due to their scarce abundance in the medium, the charm-quark annihilation rate can be considered negligible within the QGP lifetime~\cite{Braun-Munzinger:2007fth, Bodeker:2012gs}. Thus, the initial production effectively fixes the number of $c\bar{c}$ pairs. Charm quarks undergo all the evolution stages of a heavy-ion collision, and their propagation in the QGP is often modeled as Brownian motion, since, due to their large mass, they typically participate in collisions with small momentum transfer. Over the past years, numerous transport models have been developed based on this approach, addressing various aspects of heavy-quark in-medium dynamics (for recent reviews, see~\cite{Prino:2016cni,Cao:2018ews}).
Recent experimental evidence on the positive anisotropic flow of open and hidden charm hadrons~\cite{ALICE:2020iug, ALICE:2020pvw, CMS:2017vhp, ALICE:2022wpn, CMS:2024krd}, together with theoretical and phenomenological studies~\cite{Andronic:2021erx, Andronic:2023ioz,Sambataro:2025obe, Altenkort:2023oms, Altenkort:2023eav}, suggested that charm quarks could approach thermal equilibrium within the QCD medium. This supported the applicability of fluid-dynamic descriptions, and a fluid-dynamic framework for charm quark dynamics in the QGP was developed in our previous works~\cite{Capellino:2022nvf, Capellino:2023cxe}.

Assuming proximity to local thermal equilibrium, we applied in Ref.~\cite{Capellino:2022nvf} a simplified version of the method of moments~\cite{Denicol:2012cn} to compute out-of-equilibrium corrections to the charm-quark distribution function during the QGP evolution. This approach, however, does not provide a straightforward parametrization of the corresponding corrections to charm-hadron distribution functions on the freeze-out hypersurface, which we have neglected until now.

Out-of-equilibrium corrections at freeze-out are crucial for accurately describing charm-hadron yields and momentum spectra. Neglecting them led to an unphysical dependence of charm multiplicities on the heavy-quark spatial diffusion coefficient $D_s$. The $D_s$ coefficient characterizes the low-momentum interaction strength of heavy quarks in the medium, and it encodes the spatial broadening of the heavy quarks. As a result, our previous work~\cite{Capellino:2023cxe} did not allow for a systematic study of charm-hadron spectra or for the constraint of the value of the $D_s$ coefficient. In addition, the initial charm-quark distribution function was assumed to be in local thermal equilibrium, neglecting the influence of the out-of-equilibrium corrections at the initial moment of the collision.
Motivated by these considerations, in the present work, we compute the charm out-of-equilibrium corrections both at the initial stage and at the freeze-out, consistently accounting for them throughout the heavy-ion collision evolution. 
This advancement will enable future constraints on the charm diffusion coefficient through, for example, a Bayesian analysis of experimental data.

The present work is organized as follows. Sec.~\ref{sec:ooe_initial_stage} focuses on the computation of out-of-equilibrium corrections at the initial stage of the collision, and on the initialization of the charm diffusion current. In Sec.~\ref{sec:ooe_freezeout} an analytic expression of the out-of-equilibrium corrections at the freeze-out is given, employing a multi-species approach. After presenting the evolution of the charm fields in Sec.~\ref{sec:evolution_fields}, we discuss in Sec.~\ref{sec:yields_spectra} the results on the charm hadron yields and transverse momentum (\pt) spectra. Finally, Sec.~\ref{sec:discussion_ooe} focuses on the limit of the applicability of our model, exploring the upper bound of the transverse momentum up to which our approach is reliable.   
\section{Charm out-of-equilibrium corrections}
\label{sec:charm_ooe}
In kinetic theory, the state of a system is described by the single-particle distribution function $f_{\mathbf{p}}$, which encodes the phase-space density of a particle with momentum 
${\mathbf{p}}$. 
Under dissipative conditions, the distribution function is usually decomposed into a local-equilibrium contribution $f^{(0)}_{\mathbf{p}}$, and an out-of-equilibrium correction $\delta f_{\mathbf{p}}$,
\begin{equation}
    f_{\mathbf{p}} = f^{(0)}_{\mathbf{p}} + \delta f_{\mathbf{p}}.
    \label{eq:f_decomposition}
\end{equation}

This decomposition implicitly assumes that $\delta f_{\mathbf{p}}$ is a perturbative correction to the local-equilibrium distribution, such that $|\delta f_{\mathbf{p}}| \ll f^{(0)}_{\mathbf{p}}$ and higher-order corrections beyond linear order can be neglected. 
At the microscopic level, the fluid dynamic quantities can be expressed as moments of the distribution function. 
Taking as an example the density of a particle species and its associated diffusion current, in the Landau frame they can be expressed respectively as,
\begin{align}
    n_q &= \frac{1}{(2 \pi)^3} \int d^3p \ f_{\mathbf{p}} = \frac{1}{(2 \pi)^3} \int d^3p \ f^{(0)}_{\mathbf{p}}, 
    \label{eq:density_kinetic}
    \\
    \nu_q^\mu &= \frac{1}{(2 \pi)^3} \int d^3p\frac{p^{\langle \mu \rangle}}{p^0} f_{\mathbf{p}} = \frac{1}{(2 \pi)^3} \int d^3p\frac{p^{\langle \mu \rangle}}{p^0}  \delta f_{\mathbf{p}},
    \label{eq:diffusion_kinetic}
\end{align}
where the second equalities in both equations hold due to Landau matching conditions.
Notice that in the rest frame of the fluid $p^0$ coincides with the particle energy, obeying the on-shell condition $p^0 \equiv E_{\mathbf{p}}  = \sqrt{\mathbf{p}^2 + M^2}$, for a particle of mass $M$. Moreover, here the notation $p^{\langle \mu \rangle} = \Delta^{\mu \nu}p_\nu$ is employed to represent the projection of the vector $p^\mu$ onto the subspace orthogonal to the fluid four-velocity $u^\mu$, and the projection operator is defined as $\Delta^{\mu \nu} = g^{\mu \nu} + u^\mu u^\nu$.

In fluid dynamics, the deviation from local thermal equilibrium is macroscopically quantified by considering the ratio between the dissipative quantities and the equilibrium ones, defined as the inverse of \emph{Reynolds number}. In the case of diffusion processes, the inverse of the Reynolds number reads,
\begin{equation}
    \mathrm{R_\nu^{-1}} \equiv \frac{|\nu_q^\mu|}{n_q}.
    \label{eq:reynold}
\end{equation}
Since the non-equilibrium moments are defined as integrals over $\delta f_{\mathbf{p}}$, while the equilibrium ones are obtained from the equilibrium distribution, Eq.~\eqref{eq:reynold} provides a quantitative measure of the deviations from local equilibrium.

While the equilibrium distribution can be parametrized using a Maxwell-Jüttner distribution, 
in the following the calculation of the out-of-equilibrium corrections to the charm distribution function, both at the initial stage of the collision and at freeze-out, is discussed.

\subsection{Charm out-of-equilibrium corrections: initial stage}
\label{sec:ooe_initial_stage}
In the present section, we introduce an approach for modeling charm quark out-of-equilibrium corrections at the initial stage of the collision, with particular emphasis on the initialization of the diffusion current.

In our previous work~\cite{Capellino:2024zfk}, the density at midrapidity of charm–anticharm pairs produced in hard scattering processes was expressed as,
\begin{equation}
n_{q,\rm hard}(\tau_0,x,y,{\eta_s} = 0) = \frac{1}{\tau_0 \sigma^{\rm in}}n_{\rm coll}(x,y) \int dp_x dp_y \left. \frac{d^3\sigma^{Q\bar{Q}}}{dp_x dp_y d\eta_p} \right\vert_{\eta_p =0}
\label{eq:nhard_initial}
\end{equation}
where $\tau_0$ represents the fluid-dynamic initialization time, $\sigma^{\rm in}$ is the inelastic proton-proton cross section and $\left. d\sigma^{Q\bar{Q}}/dp_x dp_y d\eta_p\right\vert_{{\eta_p = 0}}$ is the charm-anticharm production cross section per unit of transverse momentum and rapidity at midrapidity computed with perturbative-QCD (pQCD) calculations from FONLL ()~\cite{Frixione:2001pa}.
In the present work, we restrict our analysis to charm quarks produced in the midrapidity region $|\eta_p|<0.5$ and assume the fluid-dynamic description to be invariant under rapidity boost.
While Eq.~\eqref{eq:nhard_initial} is valid for any system in relativistic heavy-ion collisions, this work focuses on \PbPb\ collisions at \fivenn\ within the 0--10\% centrality class. Thus, we consider $\sigma^{\mathrm{in}} = 70~\mathrm{mb}$~\cite{ALICE:2012xs} and adopt the pQCD prediction using CTEQ6.6 Parton Distribution Function (PDF)~\cite{Nadolsky:2008zw}, which yields a central value of $\left. d\sigma^{Q\bar{Q}}/d\eta_p \right\vert_{\eta_p = 0} = 0.46~\mathrm{mb}$. 
It was shown in Refs.~\cite{ALICE:2021dhb, ALICE:2023sgl} that the charm production cross section measured in pp collisions at the LHC systematically lies on the upper edge of the FONLL calculations across several center-of-mass energies. Similar deviations are expected in \PbPb collisions, implying that the use of the FONLL cross section in the present calculation may lead to an underestimation of the number of produced $c\bar{c}$ pairs. The choice of PDF and the possible nuclear modifications further affect the charm production cross section and, consequently, the initial charm yield. In the absence of a direct measurement of the charm cross section in \PbPb collisions, however, the FONLL calculation is adopted.
We underline that the purpose of this section is to validate our parametrization of the out-of-equilibrium corrections at the initial stage and at the freeze-out. For this reason, nuclear effects on the charm production cross section do not need to be included at this stage. A more detailed discussion on the cross-section, and a careful consideration of these effects, is reported in Sec.~\ref{sec:comparision_with_data}, together with the comparison with experimental data.

The density of binary collisions $n_{\rm coll}(r)$ in the transverse plane is computed with the \trento model~\cite{Moreland:2014oya} and depends on the impact parameter of the collision. 
The initial time $\tau_0$ accounts for the pure longitudinal expansion at early times, commonly referred to as \textit{Bjorken flow}~\cite{Bjorken_1983}\footnote{Notice that since we are considering an initial Bjorken flow, the space-time rapidity $\eta$ coincides with the momentum rapidity $\eta_p$. Therefore, the production cross section will be expressed in the following per unit of momentum rapidity. For notational convenience, we will drop in what follows the explicit indication of midrapidity. Notice, however, that the production cross sections are all calculated at $\eta_p = 0$.}. 

In this work, we extend Eq.~\eqref{eq:nhard_initial}, introducing a free-streaming phase between the charm production time and the hydrodynamic initialization time\footnote{In heavy-ion collisions, the earliest stage is described within the Color Glass Condensate (CGC) framework \cite{Gelis:2010nm}. Although free-streaming provides a simplified treatment of this stage, recent studies indicate that glasma effects on charm-quark transport are moderate, leading to only limited modifications of charm observables~\cite{Avramescu:2024xts, Singh:2025duj}, supporting the use of free-streaming as a reasonable approximation for early-time charm dynamics.}.
In this approach, no free-streaming stage is applied to the bulk medium, whose initial temperature profile and fluid velocity will be specified in the next section.
In the free-streaming scenario, where particles travel in straight-line trajectories, kinetic equations can be solved exactly by using the method of characteristics~\cite{romatschke_2019} and the expression for the distribution function in terms of the free-streamed coordinates is known analytically.
Applying free-streaming to the expression in Eq.~\eqref{eq:nhard_initial} leaves the physical quantities depending on the momentum, such as the production cross section, unaffected, while it influences those that depend on the spatial coordinates, as the density of binary collisions. Thus, the density of charm quarks produced in hard collisions modifies as,  
\begin{equation}
n_{q,\rm hard}(\tau_0,x,y,\eta_s = 0) = \frac{1}{\tau_0 \sigma^{\rm in}} \int dp_x dp_y n_{\rm coll}\left(x-\frac{\tau_0 p_x p^\tau}{p_T^2 + M^2}, y -\frac{\tau_0 p_y p^\tau}{p_T^2 + M^2}\right) \left. \frac{d^3\sigma^{Q\bar{Q}}}{dp_x dp_y d\eta_p} \right\vert_{\eta_p =0}, 
\label{eq:initial_density}
\end{equation}
where $p^\tau$ is the time-component of the four momentum in Milne coordinates.
By employing Landau matching conditions, and imposing that the density of charm quarks produced in the hard scattering processes is fixed to be equal to the thermal equilibrium density reported in Eq.~\eqref{eq:density_kinetic}, one can fix the integral over $p_z$ of the total charm distribution function in terms of the differential cross section,
\begin{equation}
    \int dp_z f_{\mathbf{p}} = \frac{(2 \pi)^3}{\tau_0 \sigma^{\rm in}} n_{\rm coll}\left(x-\frac{\tau_0 p_x p^\tau}{p_T^2 + M^2}, y -\frac{\tau_0 p_y p^\tau}{p_T^2 + M^2}\right) \left. \frac{d^3 \sigma^{Q\bar{Q}}}{dp_x dp_yd \eta_p} \right \vert_{\eta_p = 0} 
    \label{integral_diffusion}
\end{equation}
Thus, after the free-streaming phase, at time $\tau_0$, the charm-quark diffusion current can be expressed as the first moment of the charm distribution function, fixed in Eq.~\eqref{integral_diffusion},
\begin{align}
    \nu_q^\mu(\tau_0,x,y,\eta_s = 0) & = \frac{1}{\tau_0 \sigma^{\rm in}} \int  dp_x dp_y  \frac{p^{\langle \mu \rangle}}{p^0} n_{\rm coll}\left(x-\frac{\tau_0 p_x p^\tau}{p_T^2 + M^2}, y -\frac{\tau_0 p_y p^\tau}{p_T^2 + M^2}\right)  \left.\frac{d^3\sigma^{Q\bar{Q}}}{dp_x dp_y d\eta_p}\right\vert_{\eta_p = 0},
    \label{eq:nux_expression_fs}
\end{align}
We highlight that for symmetry reasons the diffusion current is zero at the formation time of heavy quarks. However, the dependence of $n_{\rm coll}$ on $p_x$ and $p_y$, due to the free-streaming, yields a non-zero diffusion current at time $\tau_0$ after the free-streaming phase. 

\begin{figure}
    \centering
    \subfloat
    {\includegraphics[width=.49\textwidth]{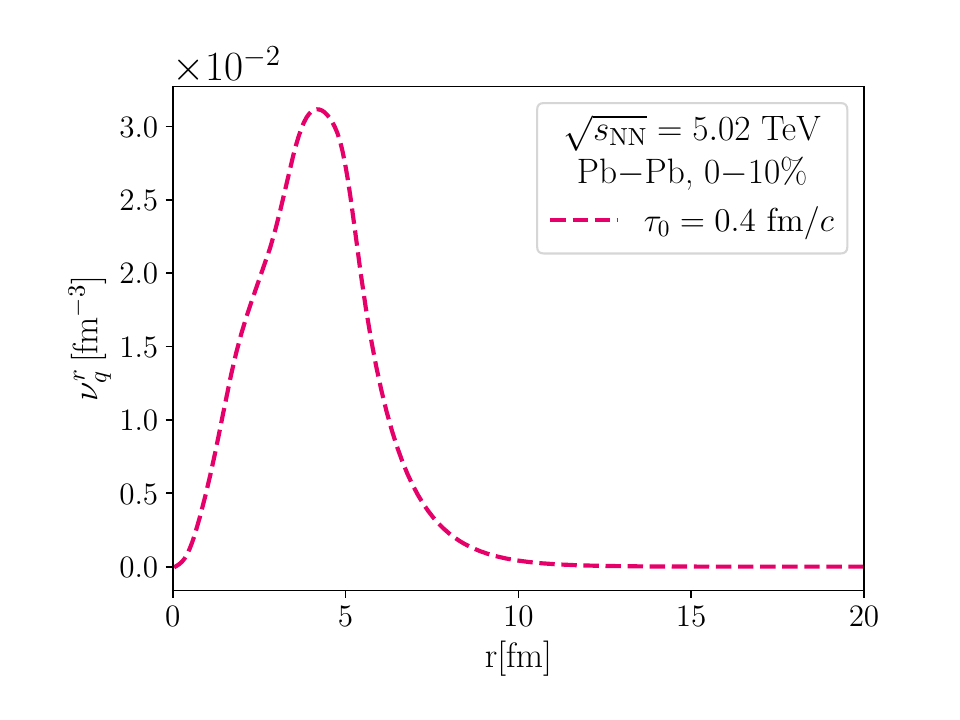}
    \label{fig:initial_nur}} \quad
    \subfloat
    {
    {\includegraphics[width=.45\textwidth]{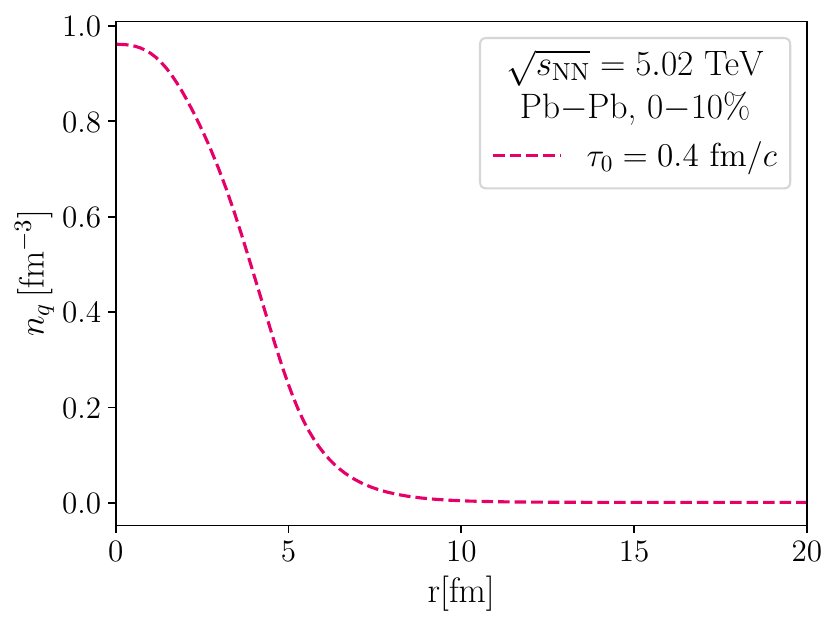}}
        \label{fig:initial_density}
    }
    \caption{Radial diffusion current (left) and charm quark density profile (right) as a function of radius at the QGP initial time $\tau_0$ for \PbPb collisions at \fivenn for the 0--10\% centrality class.}
    \label{fig:initial_charm_fields}
\end{figure}

We initialize our fields on a hypersurface of radius 30 fm and at constant proper time $\tau_0 = \mathrm{const} = 0.4 \ \mathrm{fm}/c$. Figure~\ref{fig:initial_charm_fields} shows the radial component of the diffusion current (left panel) and of the charm density (right panel) after free-streaming at $\tau_0$, considering the 0--10\% central \PbPb collisions at \fivenn. The initial production time of charm quarks is set according to the timescale for producing a charm–anticharm pair, as estimated from the uncertainty principle.
After free-streaming, the charm quarks develop a non-zero diffusion current, whose amplitude is one order of magnitude bigger than the one produced at late times $\tau \sim 10 \ \mathrm{fm}/c$, as reported in our previous work~\cite{Capellino:2023cxe}. The influence of the initial diffusion on the evolution of $\nu^\mu_q$ is investigated in detail in Sec.~\ref{sec:evolution_fields}. The density of charm quarks peaks at the center of the fireball and decreases smoothly with distance, until at $r \geq 8 \ \rm{fm}$ the number of produced charm quarks becomes negligible. Notice that after the free-streaming phase, the density profile decreases due to the Bjorken expansion along the longitudinal axis. Nevertheless, the number of charm quarks remains conserved in time due to the buildup of the diffusion current. 

\subsection{Charm out-of-equilibrium corrections: freeze-out}
\label{sec:ooe_freezeout}
In the current section, the derivation of the out-of-equilibrium corrections for charm hadrons at freeze-out, and the computation of transport coefficients for each charm-hadron species, are presented. For this purpose, a multi-species approach, as introduced in Ref.~\cite{Fotakis:2022usk}, is employed, and a fluid composed of all the charm hadrons in the Hadron Resonance Gas of charm (HRGc), interacting with the light components of the QGP, is considered.
This is a valid description of the QCD medium when the system is close to the pseudo-critical temperature, and the partonic degrees of freedom convert into the hadronic ones. However, in the absence of a charm Equation of State (EoS) that smoothly interpolates between the two regimes, the charm hadronic description is employed for the whole temperature range of the system evolution. 

The thermodynamic quantities of the mixture, as the charm-quark number density and the charm diffusion current, are expressed as a summation over the species $i$ in the HRGc, weighted by their net charm charge $q_i$,
\begin{equation}
    n_q = \sum_{i \in \mathrm{HRGc}} q_i n_i, \quad \nu^\mu_q = \sum_{i \in \mathrm{HRGc}} q_i \nu^\mu_i.
    \label{eq:total_n_and_nu}
\end{equation}
In the following, all the quantities characterized by the index $i$ refer to the hadron species $i$, while the fields denoted as $n_q$ and $\nu_q^\mu$ are associated with the total-charm dynamics.
\subsubsection{The Fokker-Planck Equation}
Assuming small momentum exchange between charm hadrons and the light partons of the medium, and considering an isotropic medium, the dynamics of charm hadrons reduces to the well-known Fokker-Planck equation, 
\begin{equation}
    p^\mu \partial_{\mu} f_{i,\mathbf{p}} = p^0 \frac{\partial}{\partial p^l} \left\{A_i p^l f_{i,\mathbf{p}} + \delta^{ml}  D_i \frac{\partial}{\partial p^m} f_{i, \mathbf{p}}\right\},
    \label{eq:FP_indep_momentum}
\end{equation}
where $l, m = 1,2,3$ represent the three spatial components. The coefficients $A_i$ and $D_i$, called respectively \emph{drag} and \emph{momentum-diffusion} coefficients, encode the interaction of the $i$-th charm hadron with the light components, and are related by the Einstein fluctuation-dissipation (EFD) relation~\cite{Svetitsky:1987gq},
\begin{equation}
    A_i = \frac{D_i}{T E_{\mathbf{p}}}.
    \label{eq:EFD}
\end{equation}
In Eq.~\eqref{eq:FP_indep_momentum}, $A_i$ and $D_i$ are assumed for simplicity to be independent of the charm-hadron momentum, similarly to Ref.~\cite{Capellino:2022nvf}. 

The spatial diffusion coefficient $D_s$ is identified as the asymptotic mean squared displacement of an ensemble of particles initially placed at the origin of the coordinate system.
In the case of charm quark dynamics in the QGP,  $D_s$ characterizes the long-wavelength properties of the transport of the charm-quark number through QCD matter, and, as long as the dynamics is non-relativistic, it can be expressed as,
\begin{equation}
D_s = \frac{T}{A(p = 0) M},
\label{eq:Ds_friction_coeff}
\end{equation}
where $M$ represents the mass of the charm quark, while $A$ is its respective friction coefficient. 
Being able to capture the charm-quark coupling with the medium at low transverse momenta, the spatial diffusion coefficient is of particular interest in phenomenological studies. In the current study, the diffusion coefficient is considered to be the same for charm quarks and each charm hadron species, and Eq.~\eqref{eq:Ds_friction_coeff} to hold in both cases. This choice reflects the idea that hadronic transport properties inherit the underlying charm–medium coupling. 
\subsubsection{Charm hadron transport coefficients}
\label{sec:charm_transport_coeff}
Having expressed the Fokker-Planck equation for each hadron species, we now consider successive moments of Eq.~\eqref{eq:FP_indep_momentum}. The $0^{\rm th}$ moment encodes the continuity equation for the hadron $i$. Taking the first moment, an equation of motion for the diffusion current of each charm hadron $\nu^\mu_{i}$ naturally arises, 
\begin{equation}
   \frac{D_s}{T}  \frac{I_{i,31}}{P_{i}} \dot{\nu}_i^\mu + \nu_i^\mu = q_i n_i D_s \nabla^\mu \alpha_{q}
   \label{eq:nui_eom_kinetic}
\end{equation}
where $I_{i,31} = \frac{1}{3}\int {d K} \ k^0 k^2  f^{(0)}_{i,\mathbf{p}}$ represents a thermodynamic integral as defined in Ref.~\cite{Denicol:2012cn}. Equation~\eqref{eq:nui_eom_kinetic} is a relaxation-type equation whose relaxation time and transport coefficient are expressed, respectively, by,
\begin{align}
    \tau_i &= \frac{D_s}{T}  \frac{I_{i,31}}{P_{i}},
    \label{eq:taui_expression}
    \\
    \kappa_i &= q_i n_{i} D_s.
    \label{eq:kappai_expression}
\end{align}
In the multi-species setup, where the diffusion of each charm hadron is tuned by $\kappa_i$ and $\tau_i$, the total-charm diffusion current $\nu_q^\mu$ obeys the following relaxation-type equation (the full calculation is reported in Appendix~\ref{app:relation_rho_nu}),
\begin{equation}
 \frac{D_s}{T} \sum_{i \in \mathrm{HRGc}} \frac{q^2_i  n_{i}}{\sum_j q^2_j n_{j}}  \frac{I_{i,31}}{P_{i}}   \dot{\nu}_q^\mu + \nu^\mu_q = D_s \sum_{i \in \mathrm{HRGc}} q^2_i n_{i}  \cdot \nabla^\mu \alpha_q.
 \label{eq:eom_total_nu}
\end{equation}
where we identify the total charm relaxation time and the total diffusion coefficient as,
\begin{align}
    \tau_q &= \frac{D_s}{T} \sum_{i \in \rm HRGc} \frac{q^2_i  n_{i}}{\sum_j q^2_j n_{j}}  \frac{I_{i,31}}{P_{i}} ,  
    \label{eq:total_tau_expression} 
    \\
    \kappa_q &= D_s \sum_{i \in \rm HRGc} q^2_i n_{i}. 
    \label{eq:total_kappa_expression}
\end{align}
Setting $i = j = 1$, the expression of $\tau_q$ 
matches the one obtained considering the total-charm diffusion as reported in Ref.~\cite{Capellino:2022nvf}. However, this does not apply to the expression of $\kappa_q$, which in the present case includes an additional charge factor, set to one in Ref.~\cite{Capellino:2022nvf}.

\subsubsection{Charm-hadrons out-of-equilibrium corrections}
\label{sec:charm_hadron_ooe}
To study the out-of-equilibrium corrections of charm hadrons at freeze-out, and following Ref.~\cite{Fotakis:2022usk}, the irreducible moment of tensor rank $l$ for each hadron species is defined as a moment of the out-of-equilibrium distribution function, 
\begin{equation}
    \rho_{i}^{ \mu_1 ... \mu_l} \equiv \Delta_{\nu_1...\nu_l}^{\mu_1...\mu_l} \int dP p^{\nu_1}...p^{\nu_l} \delta f_{i,\mathbf{p}}.
    \label{eq:irreducible_moment}
\end{equation}
It can be shown that the out-of-equilibrium corrections can be expanded as a series of irreducible moments with coefficients $a_{{i}}^{(l)}$,  
\begin{equation}
    \delta f_{i,\mathbf{p}} = f^{(0)}_{i,\mathbf{p}}  \left (\sum_{l = 0}^{\infty} a_{{i}}^{(l)} \rho_{i}^{\mu_1 ... \mu_l} p_{\langle \mu_1} ... p_{\mu_l \rangle} \right),
    \label{eq:delta_f_for_species_i}
\end{equation}
where the notation $p_{\langle \mu_1} ... p_{\mu_l \rangle} \equiv \Delta^{\mu_1 ... \mu_l}_{\nu_1 ... \nu_l} \ p^{\nu_1} ... p^{\nu_l}$ indicates the orthogonal component of the momentum with respect to the fluid velocity. Typically, the expansion in Eq.~\eqref{eq:delta_f_for_species_i} is truncated at rank $l = 2$, in order to account for the lowest-order irreducible moments explicitly appearing in the energy-momentum tensor $T^{\mu \nu}$ and the conserved current $N_q^\mu$. However, bulk pressure and shear-stress contributions from the charm hadrons are assumed to be negligible with respect to those associated with the light components. For this reason, as a first approximation, we consider only the term associated with the diffusion current, corresponding to $l =1$. Notice that by the definition of diffusion current in Eq.~\eqref{eq:diffusion_kinetic}, the irreducible moment of rank 1 corresponds to the diffusion current of the charm hadron of species $i$, i.e. $\rho^\mu_i \equiv \nu^\mu_i$. 
Starting from the Boltzmann equation and employing the Navier–Stokes approximation, a linear relation of first order in Knudsen number is established between the first irreducible moment $\rho_i^\mu$ and the total charm diffusion current $\nu^\mu_q$ \cite{Fotakis:2022usk}, 
\begin{equation}
    \rho^\mu_{i} = \bar{\kappa}_{i} \nu_q^\mu + \mathcal{O}(2),
    \label{eq:rho_vs_nu}
\end{equation} 
where the normalized diffusion coefficient is defined as,
\begin{equation}
    \bar{\kappa}_{i} = \frac{\kappa_{i}}{\sum_{j \in \mathrm{HRGc}} q_j \kappa_{j}}.
    \label{eq:kappa_bar}
\end{equation}
Notice that the only relevant order in the relation between $\nu^\mu_q$ and $\rho^\mu_i$ is of first order in Knudsen number, since the terms of order $\mathcal{O}(2)$ would give rise to higher-order contributions when considering the equation of motion of the diffusion current.
As a matter of fact all terms that are quadratic in the inverse Reynolds number are neglected, assuming $R_\nu^{-1} \ll 1$. This approximation is valid only in the limit of small contributions of $\nu^\mu_q$ to the total charm conserved current of $N^\mu_q$. The validity of this approximation will be discussed further in the next sections.

Adapting the calculation of Ref.~\cite{Capellino:2022nvf} to our multi-species approach, it is possible to express the expansion coefficient as $a_i^{(1)} = {1}/{P_i}$, where $P_i$ represents the pressure contribution of each charm hadron.
Thus, the out-of-equilibrium correction of the charm hadron of species $i$ reads,
\begin{equation}
    \delta f_{i,\mathbf{p}} = f^{(0)}_{i,\mathbf{p}}  \frac{1}{P_i}  \bar{\kappa}_{i} \nu_q^\mu p_{\langle \mu \rangle}.
    \label{eq:delta_f_hadron}
\end{equation}
Equation~\eqref{eq:delta_f_hadron} constitutes a key result of the present work, and represents the first attempt to parametrize the out-of-equilibrium corrections of charm hadrons within our fluid-dynamic approach. The expression of $\delta f_{i,\mathbf{p}}$ reduces to the one found in Ref.~\cite{Capellino:2022nvf} if we consider a fluid composed of a single particle species, marking the consistency between the two approaches. Notice that the out-of-equilibrium correction to each charm hadron depends on the total diffusion current of the charm quark $\nu^\mu_q$. This result represents a practical advantage, as in our framework, only the total diffusion current needs to be evolved, instead of one for each hadron species. 

The corrections expressed by Eq.~\eqref{eq:delta_f_hadron} are valid only in the limit $R^{-1}_\nu \ll 1$. Since the pressure of each hadron can be expressed in terms of its density, $P_i = n_i T$, the out-of-equilibrium corrections are proportional to the ratio of the diffusion current to the density. When $R^{-1}_\nu \sim 1$, the out-of-equilibrium corrections become comparable in magnitude to the equilibrium distribution, and the underlying assumption is no longer valid. Since the radial component of the diffusion current in our framework exhibits a negative profile (see Sec.~\ref{sec:evolution_fields}), the corrections can assume negative values; therefore, when $|\delta f_{i,\mathbf{p}}| \sim f^{(0)}_{i,\mathbf{p}}$, the total charm-hadron distribution function may become negative, losing its interpretation as a physical probability distribution. This clearly sets a limit to the applicability of our approach, which will be discussed in Sec. \ref{sec:discussion_ooe}.

\section{Evolution of charm fluid fields}
\label{sec:evolution_fields}
\begin{figure}[ht!]
    \centering
    \subfloat
    {\includegraphics[width=.45\textwidth]{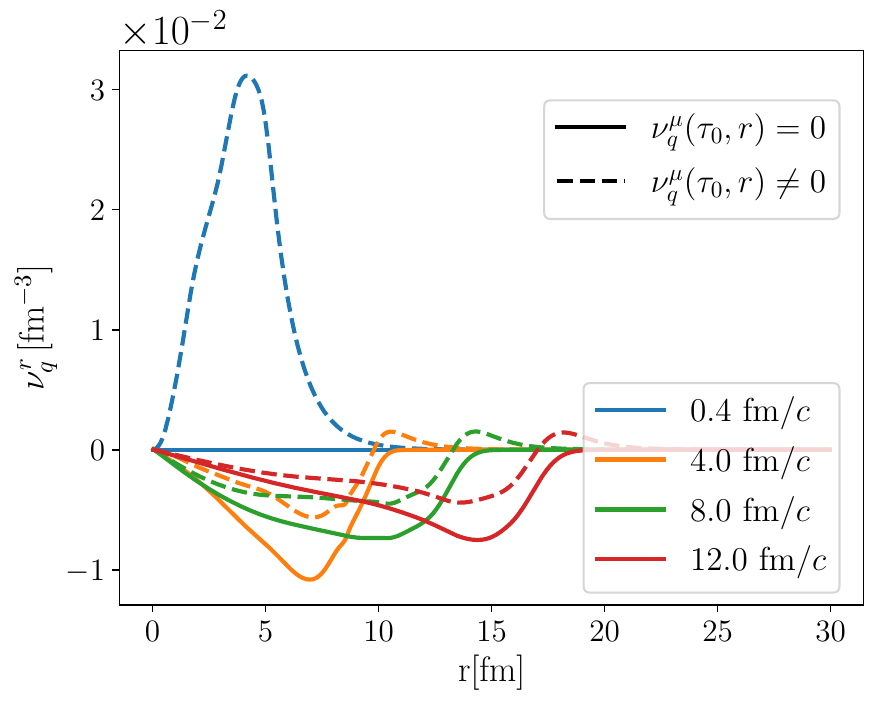}
    } \quad
    \subfloat
    {
    {\includegraphics[width=.45\textwidth]{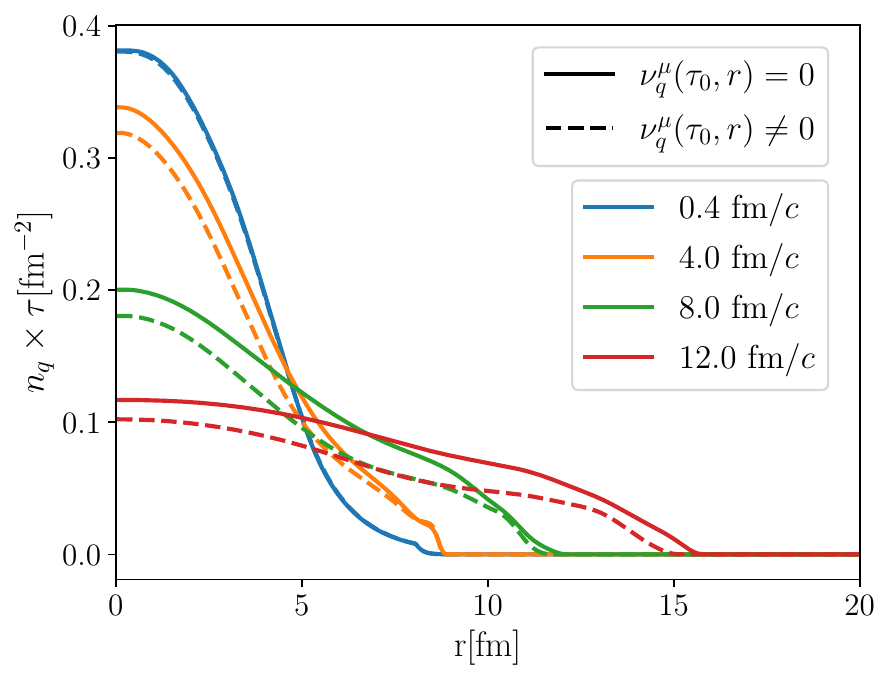}}
    }
    \caption{Radial diffusion current (left) and charm quark density profile (right) as a function of the radius for different times, fixing $D_sT = 0.24$. The comparison of a vanishing initial diffusion current case (solid lines) to that of a free-streaming initialization (dashed lines) is shown.}
    \label{fig:charm_fields_evolution}
\end{figure}

Considering the statistical azimuthal symmetry and longitudinal boost invariance of high-energy heavy-ion collisions, a $1+1 \ \mathrm{D}$ description is adopted, where the fluid fields depend only on the proper time $\tau$ and the radial coordinate $r$. Under this approximation, the fluid dynamics is governed by five independent fields: the temperature, the radial component of the fluid four-velocity, two components of the shear-stress tensor, and the bulk viscous pressure. 
The initial entropy density is computed using \trento framework~\cite{Moreland:2014oya}, where the input parameters of the collision are taken from Ref.~\cite{Vermunt:2023zsk}.
The normalization of the entropy deposition is fixed from the pion multiplicity measured by the ALICE Collaboration in \PbPb collisions at \fivenn for the 0--10\% centrality class~\cite{ALICE:2019hno}.
The initial temperature profile is obtained from the initial entropy density by inverting the EoS taken from Ref.~\cite{Floerchinger:2018pje}. In the initial state, the radial fluid four-velocity, the shear stress components, and the bulk viscous pressure are set to zero, respecting relativistic causality~\cite{Floerchinger:2017cii}.
For the fluid dynamics evolution, we employ the \fluidum 
framework~\cite{Floerchinger:2018pje} with the extension to charm fields as in Ref.~\cite{Capellino:2023cxe}. The source code employed in the current work can be found in the following GitHub repository \hyperlink{https://github.com/fafafrens/Fluidum.jl}{Fluidum.jl}. Here, the product of the diffusion coefficient with the medium temperature $D_sT$ is assumed to be constant throughout the full evolution of the fireball.  \\
\indent Figure~\ref{fig:charm_fields_evolution} shows the radial component of the diffusion current and the charm-quark density multiplied by the proper time $\tau$ as functions of the radius for different expansion times, comparing the cases of zero initial diffusion current and free-streaming initialization. The value of the spatial diffusion coefficient is fixed here to $D_sT = 0.24$, which represents the upper limit at $T=1.1 \,T_c$ reported in Ref.~\cite{Altenkort:2023oms}. In the case of free-streaming initialization, the magnitude of $\nu_q^\mu$ at later times remains systematically smaller than in the case of zero initialization, indicating that the system is closer to equilibrium. Nevertheless, the initialization does not significantly affect the evolution of the diffusion current, and it was verified that the difference between the two cases does not significantly influence the invariant yields and momentum spectra of charm hadrons. As shown in the right panel of Fig.~\ref{fig:charm_fields_evolution}, the charm-quark density is also not strongly influenced by the free-streaming initialization, and it decreases at later times due to the decreasing temperature of the fireball.  \newline
\indent Following the definition of the inverse of the Reynolds number, it is possible to compare the magnitude of the equilibrium contribution, represented by the charm quark density, to the out-of-equilibrium contribution, described by $|\nu^\mu_q|$. As reported in Appendix~\ref{app:reynolds_study}, the condition $R^{-1}_\nu \ll 1$ is not satisfied at all times and at each radius,  if one considers high values of the spatial diffusion coefficient. 
This result therefore indicates that the working assumption of small out-of-equilibrium corrections at the freeze-out may break down beyond a certain value of $D_sT$. This aspect will be investigated further in Sec.~\ref{sec:discussion_ooe}.

\section{Charm hadron yields and transverse momentum spectra}
\label{sec:yields_spectra}
The invariant spectrum of each hadron species at the freeze-out can be computed via the Cooper-Frye procedure~\cite{Cooper:1974mv},
\begin{equation}
E_\mathbf{p} \frac{d N_i}{d^3 \mathbf{p}} = \frac{g_i}{(2 \pi)^3} \int_{\Sigma} f_{i,\mathbf{p}}  p^{\mu} d \Sigma_\mu,
\label{eq:Cooper:1974mv}
\end{equation}
where $g_i$ represents the degeneracy factor of particle $i$, while $\Sigma_\mu$ is the freeze-out hyper-surface. From Eq.~\eqref{eq:Cooper:1974mv}, it is evident that the correct parametrization of the hadron distribution function $f_{i,\mathbf{p}}$ at the freeze-out is of fundamental importance to correctly compute the invariant spectrum.  
In order to account for resonance decays, we employ the FastReso package~\cite{Mazeliauskas:2018irt}, while the list of resonances is taken from the PDG~\cite{ParticleDataGroup:2022pth}. The freeze-out temperature is chosen to be $T_\mathrm{fo} = 156 \ \mathrm{MeV}$~\cite{Andronic:2021erx}.
\begin{figure}
    \centering
    \subfloat
    {\includegraphics[width=.45\textwidth]{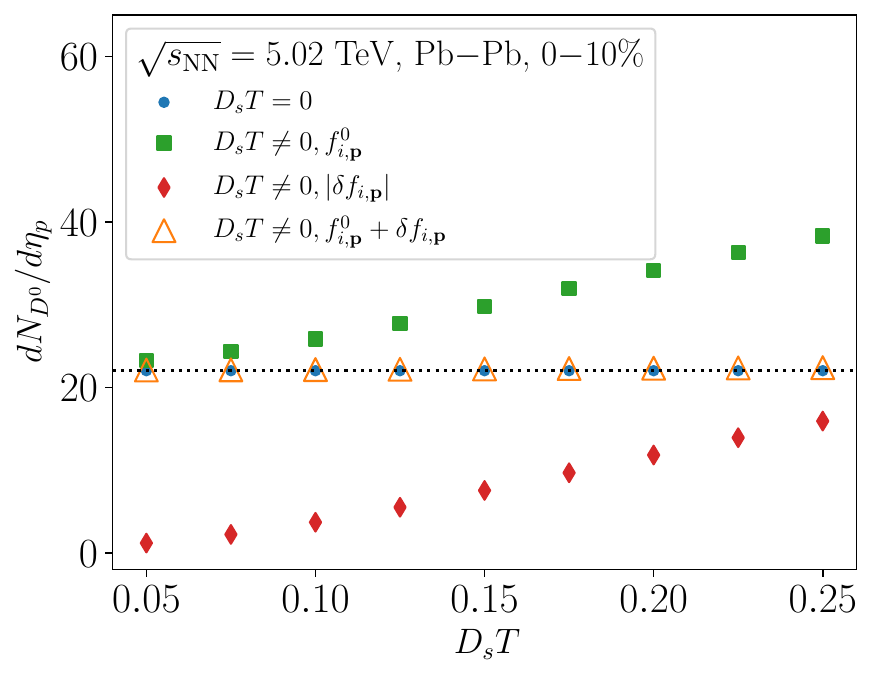}
    } \quad
    \subfloat{\raisebox{2.5mm}
    {
    {\includegraphics[width=.47\textwidth]{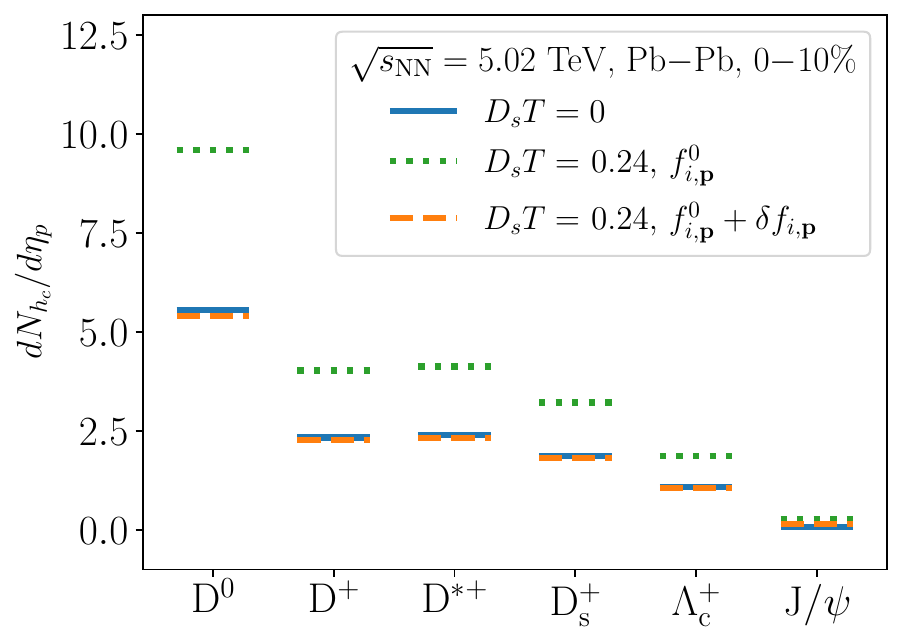}}}
    }
    \caption{Left panel: Integrated yield of $\rm D^0$ meson for different values of $D_sT$, considering a gas composed of $\rm D^0$ only. The non-diffusive case (blue markers) is compared to the equilibrium and the out-of-equilibrium contributions (green and red markers, respectively), as well as considering the total distribution function (orange markers). The $\rm D^0$ yield is also compared to the total number of charm quarks (black dashed line). Right panel: Integrated yields of various charm hadrons in the non-diffusive and in the diffusive scenarios, with and without out-of-equilibrium corrections.}
    \label{fig:invariant_yield}
\end{figure}
The yield per unit of rapidity $\frac{dN_i}{d\eta_p}$ of each charm hadron is obtained by integrating Eq.~\eqref{eq:Cooper:1974mv} in a \pt range from $0$ to $10 \ \mathrm{GeV}/c$. The conserved number of charm quarks produced in the QGP redistributes among the various charm-hadron species at freeze-out according to their thermal weights. Consequently, the integrated yield of charm hadrons is expected to be independent of the value of $D_sT$.

To validate the expression for the charm-hadron out-of-equilibrium correction in Eq.~\eqref{eq:delta_f_hadron}, a toy model is first considered, consisting of a gas composed of a single charm hadron, the $\mathrm{D^0}$ meson. Within this simplified scenario, the number of charm quarks in the QGP is expected to match the multiplicity of $\mathrm{D^0}$ mesons produced at freeze-out.
The left panel of Fig.~\ref{fig:invariant_yield} shows the invariant yield of $\mathrm{D^0}$ mesons as a function of the spatial diffusion coefficient $D_sT$, varied within the range predicted in Ref.~\cite{Altenkort:2023oms}. Comparisons are made between the non-diffusive case (blue markers) and the diffusive case, considering separately the equilibrium and the out-of-equilibrium contributions (green and red markers, respectively), as well as the total distribution function (orange markers).  The $\mathrm{D^0}$ yield is also compared to the total number of charm quarks (black dashed line) produced at the beginning of the collision.  
The spatial diffusion coefficient governs the evolution of $\nu^\mu_q$, such that higher values of $D_sT$ produce larger out-of-equilibrium corrections (see Sec.~\ref{sec:charm_transport_coeff}). Neglecting $\delta f_{i, \mathbf{p}}$ results in an unphysical dependence of the hadron multiplicity on $D_sT$. By including the out-of-equilibrium corrections, the invariant yield is comparable with both the non-diffusive case and the total number of charm quarks in the QGP across the entire range of $D_sT$. This agreement confirms the validity of the parametrization of the out-of-equilibrium corrections. Notice that the corrections contribute negatively to the total yield, counterbalancing the unphysical enhancement observed for $f^{(0)}_{i,\mathbf{p}}$ only, and are thus reported here in absolute value. The out-of-equilibrium fraction of the invariant yield does not exceed the equilibrium one throughout the whole range of $D_sT$ probed, even if their ratio increases due to the proportionality of $\delta f_{i,\mathbf{p}}$ on the spatial diffusion coefficient. Notice that the $\delta f_{i,\mathbf{p}}$ linearly depends on the transverse momentum. 
However, given that the largest fraction of the invariant yield arises from the low-\pt region, we observe that the contribution to the yield from the equilibrium part of the distribution function exceeds the out-of-equilibrium one for any investigated value of $D_sT$.
For the largest value of $D_sT$ tested, i.e. $D_sT=0.24$, the ratio between the contributions reaches a value of around $40 \%$.

A gas composed of all hadrons in the HRGc is further considered, and the right panel of Fig.~\ref{fig:invariant_yield} shows the multiplicity of the most abundant charm-hadron species for $D_sT = 0.24$, including contributions from all resonance decays. Comparisons are made between the non-diffusive case (blue solid line) and the diffusive case, both with (orange dashed line) and without (green dotted line) out-of-equilibrium corrections. The inclusion of out-of-equilibrium corrections makes the multiplicity compatible with the one obtained in the non-diffusive case for all the charm hadrons considered.

\begin{figure}
    \centering
    \subfloat
    {\includegraphics[width=.45\textwidth]{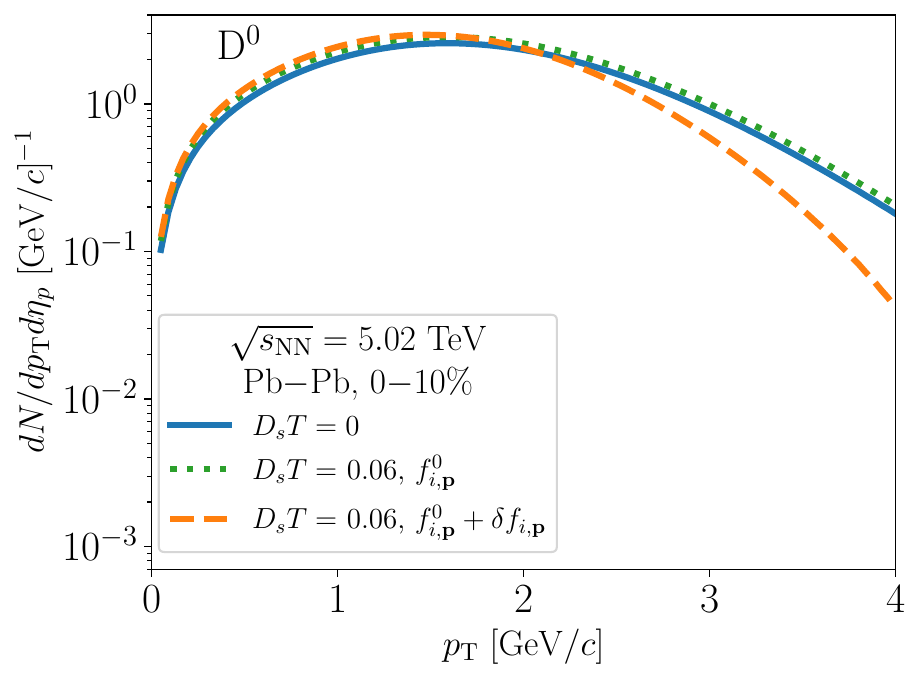}
    } \quad
    \subfloat
    {
    {\includegraphics[width=.45\textwidth]{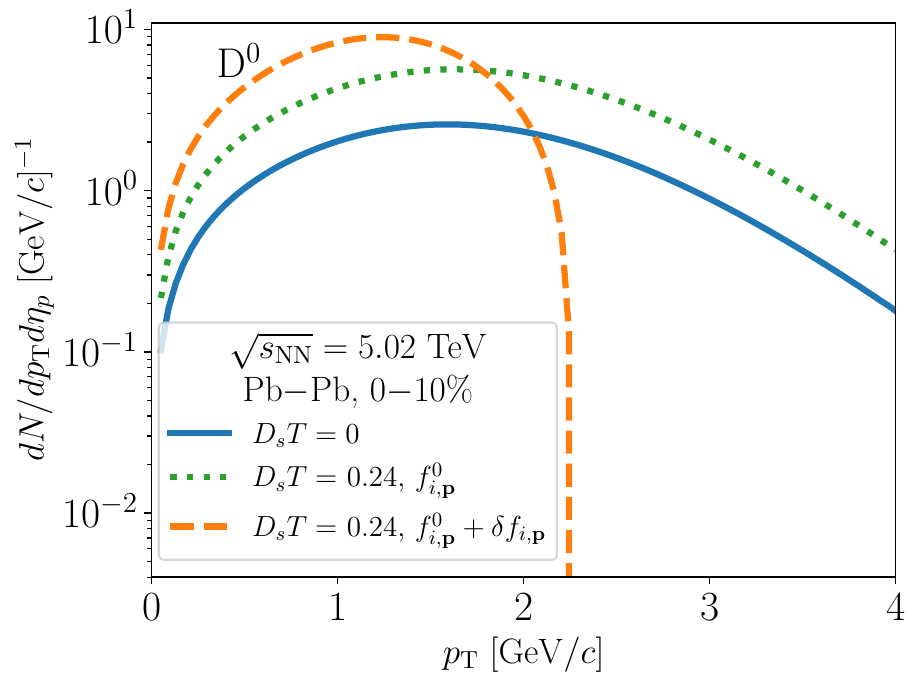}}
    }
    \caption{Momentum distribution of $\mathrm{D^0}$ mesons for $D_sT = 0.06$ (left) and $D_sT = 0.24$ (right), comparing the non-diffusive case (blue line) with the diffusive scenario, shown for the equilibrium (green line) and total (orange line) distribution functions. 
    }
    \label{fig:D0_momentum_distribution}
\end{figure}

The parametrization of the out-of-equilibrium corrections enables the study of charm-hadron \pt spectra for different values of $D_sT$. Figure~\ref{fig:D0_momentum_distribution} presents the charm-hadron invariant spectrum for the non-diffusive case (blue solid line) and for $D_sT = 0.06$ (left panel) and $D_sT = 0.24$ (right panel), both with (orange dashed line) and without (green dotted line) out-of-equilibrium corrections. The value of the spatial diffusion coefficient clearly influences the shape of the invariant spectrum while keeping the charm-hadron multiplicity constant. We notice here that the out-of-equilibrium corrections are positive for \pt $\lesssim 2 \ \mathrm{GeV}/c$ for both values of $D_sT$ considered.
For $D_sT = 0.24$, the momentum spectrum becomes negative at around $p_{\rm T} \sim 2 \ \mathrm{GeV}/c$. This behavior occurs because in this \pt region, the out-of-equilibrium corrections exceed the equilibrium contribution in magnitude. 
Consequently, the current model provides a physically meaningful description only within a limited range of spatial diffusion coefficient values. As a first approximation, the value of $D_sT$ in the current model is considered to be constant throughout the whole QGP evolution. Assuming a linear dependence of $D_sT$ on the temperature, consistent with the behavior found in lQCD calculations~\cite{Altenkort:2023oms}, leads to a reduction of the momentum range of applicability by about 5\% at high $p_{\rm T}$.
Other approaches that maintain a positive distribution function across the entire considered regime, such as the maximum entropy approach~\cite{Chattopadhyay:2023hpd}, will be explored in future works. 

\section{Comparison of the model with the available experimental data}
\label{sec:comparision_with_data}
\begin{figure}
    \centering
    \subfloat
    {\includegraphics[width=.48\textwidth]{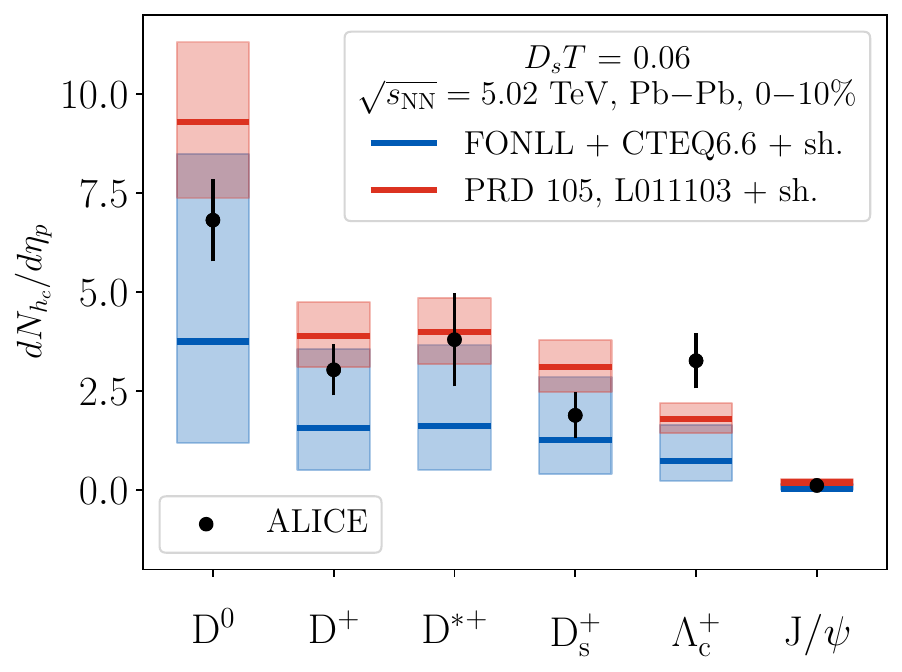}
    \label{fig:experimental_mult}} \quad
    \subfloat
    {
    {\includegraphics[width=.48\textwidth]{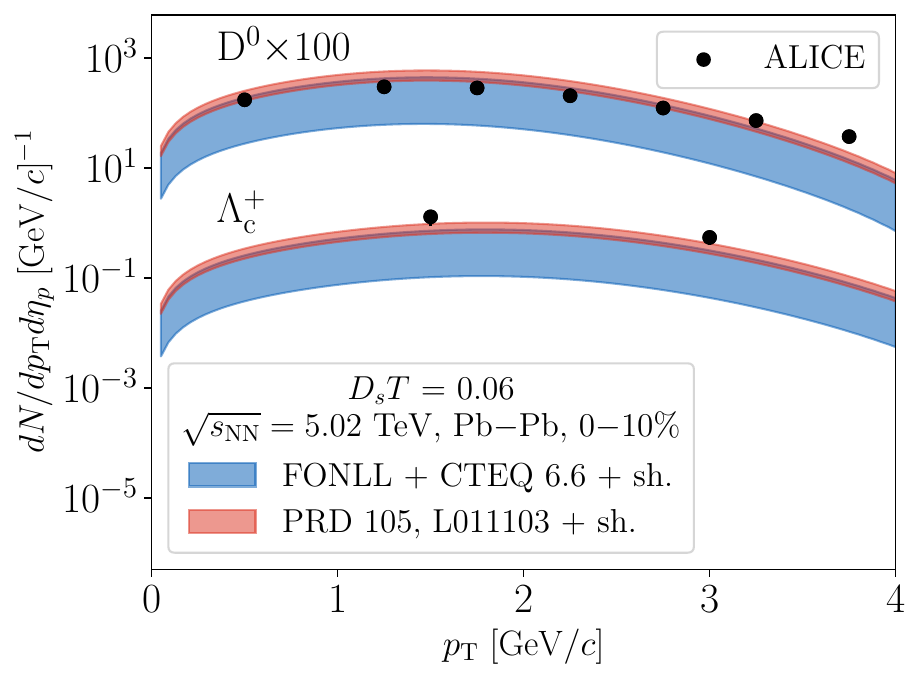}}
        \label{fig:experimental_spectrum}
    }
    \caption{Integrated yield for various charm hadronic species (left) and the $\mathrm{D^0}$ and $\mathrm{\Lambda_c^+}$ \pt spectra (right), considering both the FONLL charm production cross section~\cite{Cacciari:2012ny} and the cross section measured by ALICE multiplied~\cite{ALICE:2021dhb} by a correction factor accounting for shadowing, compared with the available experimental data.}
    \label{fig:yields_experimental}
\end{figure}

\begin{figure}
    \centering
    {\includegraphics[width=.48\textwidth]{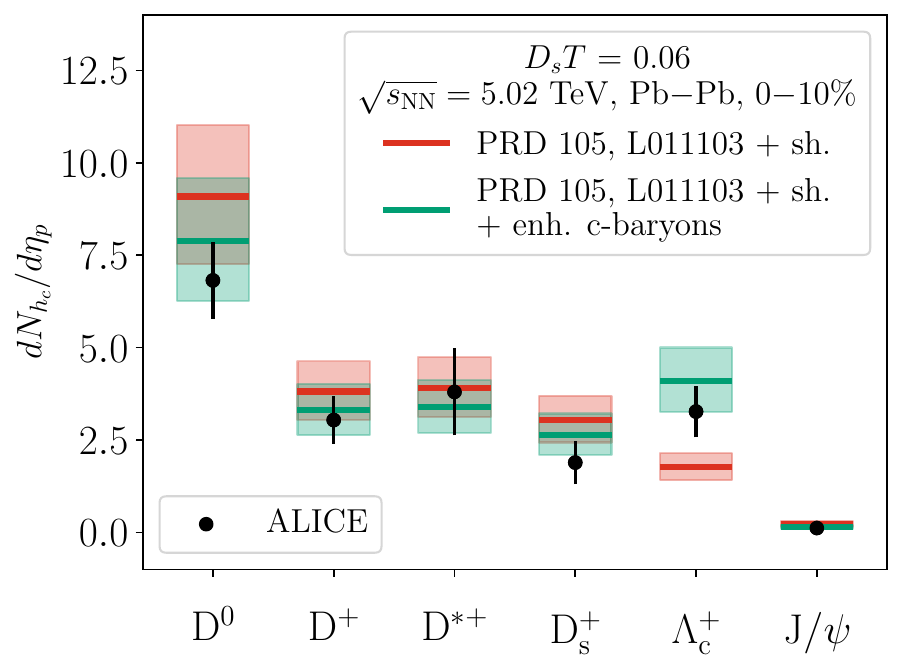}}
    {\includegraphics[width=.48\textwidth]{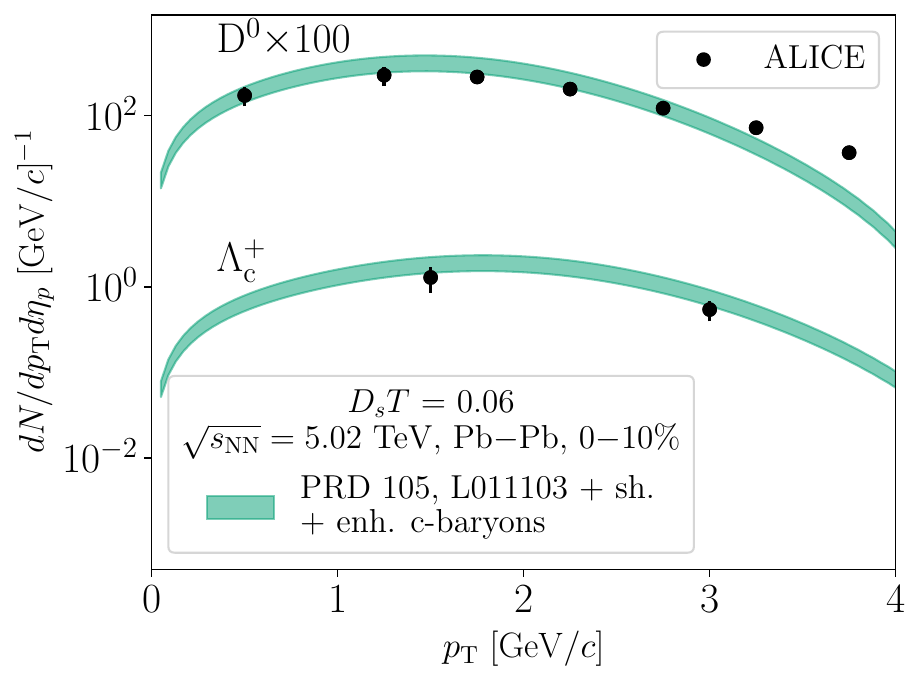}}
    \caption{Left: integrated yield for various charm-hadronic species, considering as charm production cross section the one measured by ALICE~\cite{ALICE:2021dhb}, with and without charm-baryon enhancement (see text for details). Right: the \pt spectrum of $\mathrm{D^0}$ and $\Lambda_{\rm c}^+$ including charm-baryon enhancement, in comparison with experimental measurements from ALICE.}
    \label{fig:baryon_enh}
\end{figure}

The charm-hadron yield is directly proportional to the charm production cross section. However, experimental measurements of the total charm cross section in \PbPb collisions at LHC energies are currently not available. Therefore, in our computation we employ pQCD calculations from FONLL~\cite{Cacciari:2012ny} and experimental measurements from pp collisions.
Considering the CTEQ6.6 as PDF model~\cite{Nadolsky:2008zw}, the FONLL calculation predicts in the case of pp collisions at \fivenn a differential charm-quark production cross section at midrapidity of $\left. d\sigma^{Q\bar{Q}}/d\eta_p \right\vert_{\eta_p = 0} = 0.46^{+0.60}_{-0.32} ~\mathrm{mb}$. The error values reported here include both the uncertainty on the PDF calculations as well as on renormalisation and charm-mass scales.\newline
\indent In pp collisions, the charm cross section at midrapidity is measured by the ALICE Collaboration~\cite{ALICE:2021dhb}, $\left. d\sigma^{Q\bar{Q}}/d\eta_p \right\vert_{\eta_p = 0} = 1.17^{+0.14}_{-0.11} ~\mathrm{mb}$. This value lies at the upper edge of the pQCD calculations~\cite{ALICE:2019nxm,ALICE:2021mgk,ALICE:2023sgl}. To account for nuclear modification effects in \PbPb, in the present section, both the FONLL and the ALICE cross sections are multiplied by a folding factor that incorporates shadowing, energy loss, and saturation. The estimate of this reduction factor is taken from Ref.~\cite{Andronic:2021erx}, and at midrapidity is equal to $0.65 \pm 0.12$. \newline
\indent In the left panel of Fig.~\ref{fig:yields_experimental}, we present a study of the integrated yields of various charm hadron species, including out-of-equilibrium corrections and assuming a diffusion coefficient $D_sT = 0.06$. 
The different colour codes distinguish the charm-production cross sections used as input in Eq.~\ref{eq:initial_density} to initialize the charm-quark fields. The blue marker corresponds to the FONLL-based calculation, whereas the red marker is obtained using the ALICE measured charm cross section. The shaded bands indicate the uncertainties associated with each choice of the charm cross section. For the FONLL calculation, the uncertainty estimate includes both PDF uncertainties as well as renormalisation and mass scale uncertainties, similarly to Ref.~\cite{ALICE:2019nxm}. For the ALICE experimental cross section, the uncertainty is obtained by adding in quadrature the statistical and systematic uncertainties reported in Ref.~\cite{ALICE:2021dhb}. In order to account for nuclear modification effects, both model calculations were further combined with the uncertainty associated with shadowing, as reported in Ref.~\cite{Andronic:2021erx}.
The reported hadron yields are compared with experimental data from the ALICE Collaboration~\cite{ALICE:2021rxa, ALICE:2023gco, ALICE:2021bib, 2022136986}. 
Our model calculations are consistent within uncertainties for all meson species, both when using the FONLL prediction and the charm production cross section measured by ALICE. However, the large theoretical uncertainties associated with the FONLL cross section limit the predictive power of the current approach for the integrated yields. A deviation between our model and the measured yield is observed in the case of the $\Lambda_c^+$ baryon: this behavior might be caused by the absence of higher resonance states in the PDG, as already underlined in Refs.~\cite{ Andronic:2021erx, He:2019vgs, HE2019117}.\\ 
\indent In the right panel of Fig.~\ref{fig:yields_experimental}, the \pt spectra of the $\mathrm{D^0}$ meson and of the $\mathrm{\Lambda^+_c}$ baryon are shown in comparison with the ALICE data~\cite{ALICE:2021rxa}. In the case of the $\mathrm{D^0}$, our fluid-dynamic model captures the spectrum shape in the case of FONLL calculations up to \pt $\approx 2-3 \ \mathrm{GeV}/c$. The different values of the charm cross section result in a different overall normalization of the spectrum, and, when employing the value from pp collisions measured by ALICE, our prediction tends to overestimate the experimental data. On the other hand, consistently with what was observed for the integrated yield, both cross sections underestimate the experimental \pt spectrum for the $\mathrm{\Lambda^+_c}$ baryon.\\
\indent Given the tension between our model and the measured yield of the $\mathrm{\Lambda^+_c}$ baryon, we apply the following method to account for the possible missing resonances feeding down to charm baryons.
Inspired by the work carried out in Ref.~\cite{Andronic:2021erx}, we increase the thermal weight of all the excited charm baryons in the PDG by a factor of 3. Notice that since the charm cross section measured in pp should comprise all the possible resonant states, its value should be kept constant when applying this procedure. \\
\indent In the left panel of Fig.~\ref{fig:baryon_enh}, we report the charm-hadron yields computed by employing the ALICE charm cross section, with and without including the charm-baryon enhancement. As expected, the increase in the number of charm baryons due to the tripled statistical weights subsequently implies a decrease in the yield of charm mesons. Considering the charm-baryon enhancement, our predictions for all the charm-hadron species are compatible with the experimental data within uncertainties. In the right panel of Fig.~\ref{fig:baryon_enh}, the \pt spectrum of $\mathrm{D^0}$ and $\mathrm{\Lambda^+_c}$ including the charm-baryon enhancement is shown in comparison with experimental measurements from ALICE. 
The experimental measurements are reproduced up to $3 \ \mathrm{GeV}/c$ in \pt for both hadron species.\\
We emphasize that the results shown in Figs.~\ref{fig:yields_experimental} and~\ref{fig:baryon_enh} are obtained for a diffusion coefficient $D_s T = 0.06$. This value corresponds to the largest diffusion coefficient for which the present model yields positive particle spectra over the full $p_{\rm T}$ range considered. The same considerations concerning the onset of negative particle spectra remain valid. The predictions presented in this section constitute a first comparison of our model with the available experimental data. A systematic analysis based on a Bayesian fit to the experimental measurements will be carried out in future work. Moreover, this procedure represents a simplified attempt to account for the missing baryon resonant states, and it should not be interpreted as a precise description, but rather as a preliminary estimate. In future studies, a resonance list that includes an augmented set of charm baryon states, given by predictions of the relativistic quark model (RQM)~\cite{Ebert:2011kk}, will be used.

\section{On the onset of charm number fluid dynamics}
\label{sec:discussion_ooe}
Negative hadron momentum spectra above a certain \pt can be expected in fluid dynamic models. As a matter of fact, fluid dynamic fields correspond to moments of the distribution function, and the reconstruction of $f_{\mathbf{p}}$ from the former is reliable only within the radius of convergence of the associated Taylor series.
In the current section, an estimate is provided for the transverse momentum up to which the charm-hadron momentum spectrum remains positive, ensuring the validity of the current approximation. 

Employing thermodynamic relations and the definition of the normalized diffusion coefficient, the total charm-hadron distribution function can be expressed as, 
\begin{equation}
     f_{i,\mathbf{p}} = f^{(0)}_{i,\mathbf{p}}
     \left[1+
     \left(\frac{1}{T} \right) \frac{q_i}{\sum_j q_j^2 n_j} \nu_q^\mu p_\mu  \right].
     \label{eq:total_f}
\end{equation}
As observed in the left panel of Fig.~\ref{fig:charm_fields_evolution}, the diffusion current exhibits negative values in the spatial domain explored by the fireball. Therefore, there can be regions on the freeze-out surface where the out-of-equilibrium corrections are negative, possibly exceeding in magnitude the equilibrium component. There can therefore be three different scenarios:
\begin{itemize}
    \item The total distribution function is positive on the entire freeze-out surface, and therefore its definition of physical probability distribution holds for any value of \pt. This automatically yields positively defined momentum distributions of charm hadrons; 
    \item The total distribution function is negative on a portion of the freeze-out surface, but the positive contribution exceeds in magnitude the negative one. Therefore, after computing the freeze-out integration, one obtains a positive value for the \pt distribution;
    \item The total distribution function is negative on a portion of the freeze-out surface, and the positive contribution is smaller in magnitude than the negative one. Therefore, after computing the freeze-out integration, one obtains a negative value for the \pt distribution.
\end{itemize}
Given that the distribution function on the freeze-out surface lives on a 6-dimensional space (it depends on the freeze-out coordinates and on the momentum coordinates), it is not trivial to analytically determine the region in which it is positive. However, it is possible to study the integrand in Eq.~\eqref{eq:Cooper:1974mv}, and determine its positivity after integrating out the space-rapidity and azimuthal angle dependence (see Appendix~\ref{app:integrand_study}). 
In the present discussion, we restrict ourselves to studying numerically in which \pt region the momentum distributions are positive, i.e., after the freeze-out integration. 
\begin{figure}
    \centering
    \subfloat
    {\includegraphics[width=.48\textwidth]{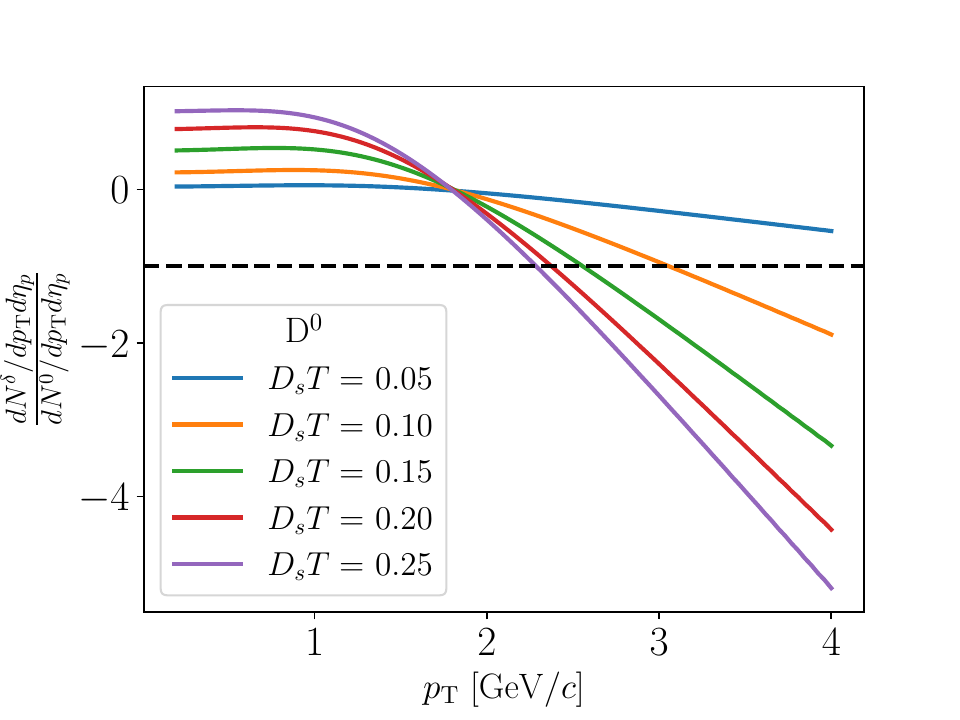}
    } \quad
    \subfloat
    {
     {\includegraphics[width=.44\textwidth]{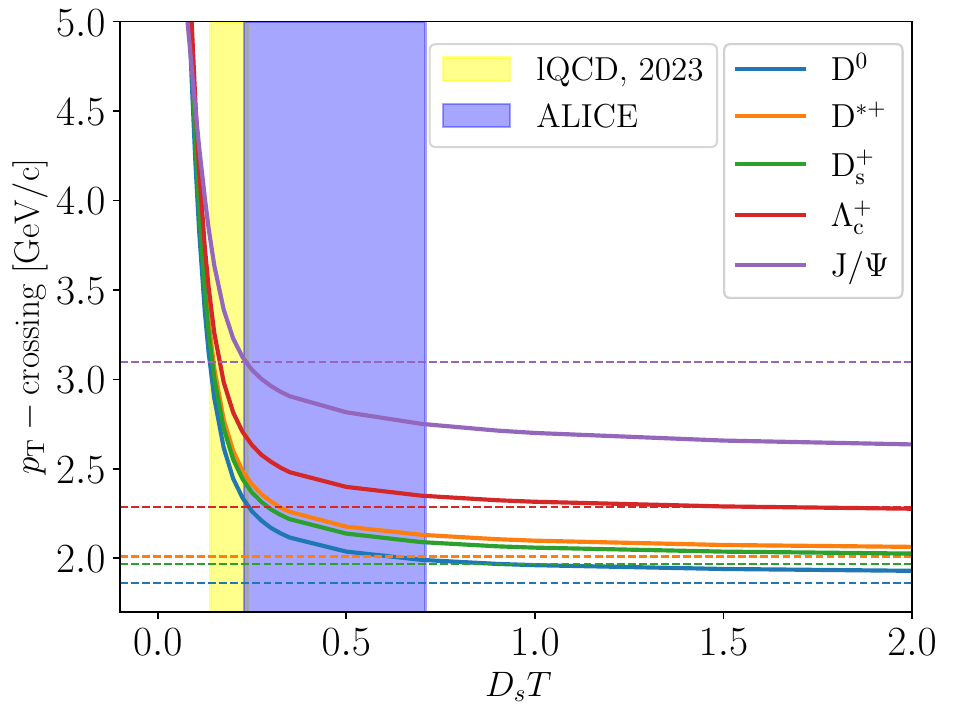}} 
     }
    \caption{Left panel: ratio between the out-of-equilibrium and the equilibrium contribution of $\rm D^0$ meson momentum spectrum for different values of $D_sT$. To guide the eye, a black dashed line has been drawn at $-1$. Right panel: Study of the zero-crossing of \pt spectra as a function of $D_sT$. At high values of $D_sT$, the value of \pt-crossing approaches the mass of the corresponding charm hadron (reported as dashed lines), with the exception of $\rm J/\Psi$ meson.}
    \label{fig:pT_crossing_study}
\end{figure}
\\ \indent
Starting from Eq.~\eqref{eq:Cooper:1974mv}, the equilibrium and out-of-equilibrium contributions to the transverse-momentum spectrum of $\mathrm{D^0}$ meson at midrapidity are respectively defined as,
\begin{equation}
    \frac{dN^{(0)}}{dp_Td\eta_p} =\frac{g_i}{(2 \pi)^3} \int_{\Sigma} f_{i,\mathbf{p}}^{(0)}  p^{\mu} d \Sigma_\mu\,,\qquad \frac{dN^{(\delta)}}{dp_Td\eta_p} = \frac{g_i}{(2 \pi)^3} \int_{\Sigma} \delta f_{i,\mathbf{p}}  p^{\mu} d \Sigma_\mu\,.
\end{equation}
In the left panel of Fig.~\ref{fig:pT_crossing_study}, the ratio between $dN^{(\delta)}/(dp_Td\eta_p)$, and $dN^{(0)}/(dp_Td\eta_p)$, is reported as a function of \pt at midrapidity for different values of $D_sT$. We observe that for \pt values approximately lower than the ${\rm D^0}$ mass ($M_{\rm D^0}\approx 1.8$ $\mathrm{GeV}/c^2$), the total spectrum is positive, since the out-of-equilibrium contribution assumes positive values. For larger values of \pt, the correction becomes negative, yielding negative spectra values once the out-of-equilibrium component exceeds the equilibrium one in magnitude. In the plot, this is shown by the \pt value at which the colored lines intercept the dashed black line, which is set to -1. We will refer to these points as \pt-crossings. Moreover, we notice that at $D_sT \sim 0.25$, the out-of-equilibrium component of the particle spectrum becomes of the same order of magnitude as the equilibrium contribution, even in the low \pt region. Therefore, for this value of the spatial diffusion coefficient, the assumption that $|\delta f_{\mathbf{p}}| \ll f^{(0)}_{\mathbf{p}}$ is not valid anymore, and the Taylor expansion of the total distribution function developed in Sec.~\ref{sec:charm_ooe} is not applicable in the whole \pt range probed.

In the right panel of Fig.~\ref{fig:pT_crossing_study}, the \pt-crossings are displayed as a function of $D_sT$, for several charm-hadronic species.
It can be observed that, as $D_sT$ increases, the \pt-crossing value decreases, indicating that the range of applicability of our approach is sensitive to the spatial diffusion coefficient. As expected, small values of $D_sT$ imply that charm hadrons are closer to equilibrium, extending the \pt regime in which the fluid dynamic description is applicable. In the case of open-charm states, at high $D_sT$ the \pt-crossing approaches the value of the mass of the respective hadron, reported as colored dashed lines.  
The only exception is $\rm J/\Psi$, which has double charm charge. In this case, one observes that the regime of applicability is limited to below its mass already at relatively low values of the spatial diffusion coefficient ($D_sT\lesssim0.25$).  
However, of course, the approach described in this work assumes that the dynamics of charm is purely governed by soft scatterings with the light partons of the medium, and it neglects the radiative processes that are dominant at high \pt. Therefore, although the fluid dynamic approach might be mathematically applicable, it would not provide a reliable description of the physics at larger transverse momenta. Additionally, it is possible to investigate the \pt region of applicability of our model, considering the ranges of $D_sT$ extracted from lattice-QCD (lQCD) calculations~\cite{Altenkort:2023oms} (here taken at $T=1.1\,T_c$), and from fits to experimental measurements of charm hadrons' spectra end elliptic flow from the ALICE collaboration~\cite{ALICE:2021rxa}. We report these ranges in Fig.~\ref{fig:pT_crossing_study} as yellow and blue bands, respectively. In the case of the upper limit predicted by lQCD calculation, our model is applicable for \pt $\lesssim 2.5 \ \mathrm{GeV}/c$ in the case of open-charm hadrons, while it extends to values of \pt $\lesssim 3 \ \mathrm{GeV}/c$ for the $\rm J/\Psi$ meson. Considering the upper bound predicted by ALICE fits, where $D_sT \sim 0.7$, the range of applicability is restricted to the \pt $\lesssim 2 \ \mathrm{GeV}/c$ for open charm hadrons, and to \pt $\lesssim 2.8 \ \mathrm{GeV}/c$ for the $\rm J/\Psi$ meson. 
\section{Summary and outlook}
This work presents an approach to compute the out-of-equilibrium corrections to the charm-quark distribution function, both during the initial stage of heavy-ion collisions and at the freeze-out.

At the initial stage, charm quarks are assumed to undergo a free-streaming phase between their production and the beginning of the fluid dynamics description. Employing Landau matching conditions and decomposing the charm production cross section into equilibrium and out-of-equilibrium components, an expression for the charm out-of-equilibrium distribution function can be determined. The free-streaming stage generates a non-vanishing diffusion current at the beginning of the hydrodynamic evolution; however, this initial contribution is found to have little impact on the subsequent evolution of the charm-quark fields, which at late times remain comparable to those obtained for $\nu^r(\tau_0,r) = 0$.\\
\indent The out-of-equilibrium corrections at freeze-out are parametrized within a multi-species approach, decomposing the charm-quark fields into the individual contributions of each charm hadron species in the HRGc. This procedure allows the derivation of the out-of-equilibrium corrections of each charm hadron. Incorporating these corrections at freeze-out enables a consistent computation of the invariant yields. In particular, it was observed that neglecting $\delta f_{i,\mathbf{p}}$ produces an unphysical dependence of the charm hadron multiplicity on $D_sT$, while including the corrections restores compatibility with the non-diffusive case across the entire $D_sT$ range explored, thereby confirming the self-consistency validity of our framework. \\
\indent In Sec.~\ref{sec:comparision_with_data}, a comparison between our fluid-dynamic approach and the experimental data is presented. Using either the FONLL prediction or the charm production cross section measured by the ALICE Collaboration, the invariant yields of charm mesons are found to be compatible with the data. However, a tension in the baryon sector, presumably due to missing high resonance states, is observed. The missing resonances can be accounted for by considering tripled statistical weights for all the excited charm baryons, similarly to what is done in Ref.~\cite{Andronic:2021erx}. By employing this procedure, and considering the cross section measured by ALICE Collaboration, our model is able to reproduce the experimental invariant yields of various hadron species, as well as their \pt spectra.\\   
\indent We underline that the present approach to compute out-of-equilibrium corrections relies on the assumption that $|\delta f_{\mathbf{p}}| \ll f^{(0)}_{\mathbf{p}}$. For $D_sT\lesssim0.1$, this is valid in the entirety of the explored \pt range. For $0.1\lesssim D_sT\lesssim0.2$, the contribution to the integrated yield from the out-of-equilibrium component is smaller than the equilibrium component; however, the applicability of the model is restricted to \pt $\lesssim 2.5-3$ GeV$/c$, where the spectrum assumes positive values. For $D_sT \gtrsim0.2 $,  $|\delta f_{\mathbf{p}}| \approx f^{(0)}_{\mathbf{p}}$ both at low and large transverse momenta, meaning that our model is not applicable anymore, and restricting its validity to lower values of the heavy-quark spatial diffusion coefficient. This behavior is also signaled by the inverse Reynolds number, as observed in Fig.~\ref{fig:nu_n_ratios} of Appendix \ref{app:reynolds_study}. The latter, in fact, grows larger than 1 for high values of $D_sT$, yielding large out-of-equilibrium corrections on the freeze-out surface. In summary, the charm-hadron momentum spectra are observed to become negative beyond a certain value of \pt and/or $D_sT$, which sets a limit on the applicability of the framework. Future developments may explore alternative methods that guarantee positivity of the distribution function across the full \pt range, such as the maximum entropy approach proposed in Ref.~\cite{Chattopadhyay:2023hpd}.
This advancement will enable future constraints on the charm diffusion coefficient and the charm production cross section, if not yet experimentally measured, through, for example, a Bayesian analysis of experimental data.

\section*{Acknowledgements}
The authors thank Stefan Floerchinger for insightful discussions, Andreas Kirchner for assistance with setting up the FastReso framework and collaborating on related subjects, Andrea Beraudo for clarifications concerning the charm quark density, Guillermo Sanchez for constructive mathematical discussions and Anton Andronic for clarifications regarding the SHMc approach to charm-baryon enhancement. This work is part of and supported by the DFG Collaborative Research Centre ``SFB 1225 (ISOQUANT)''.
Computational resources have been provided by the GSI Helmholtzzentrum f{\"u}r  Schwerionenforschung.

\bibliographystyle{apsrev4-2}
\bibliography{ref}

\appendix
\section{Equation of motion for the total-charm diffusion current}
\label{app:relation_rho_nu}
In this section, we find an expression for the equation of motion obeyed by the total-charm diffusion current $\nu^\mu_q$. As reported in Eq.~\eqref{eq:rho_vs_nu}, under Navier-Stokes approximation, there exists a linear relation between the rank-1 irreducible moment $\rho^\mu_i$ and the charm quark diffusion current,
\begin{equation}
    \rho^{\mu}_{i}  = \bar{\kappa}_{i} \nu_q^\mu + \mathcal{O}(2).
    \label{eq:rho_appendix}
\end{equation}
Even if this relation was obtained in Navier-Stokes approximation, when we exploit this expression in the equation of motion of $\rho^\mu_i$, all the $\mathcal{O}(2)$ terms will become $\mathcal{O}(3)$ terms, and thus can be neglected.\\
The equation of motion for the rank-1 irreducible moment reported in Ref.~\cite{Fotakis:2022usk} reads,
\begin{equation}
    \tau_{i} \dot{\rho}^{\langle \mu \rangle}_{i} + \rho^{\mu}_{i} = \kappa_{i} \nabla^\mu \alpha_q + \mathcal{O}(2).
    \label{eq:eom_for_rho_app}
\end{equation}
Starting from this equation, one can plug in the expression for $\nu^\mu_q$ in Eq.~\eqref{eq:rho_appendix}, multiply by the charm charge $q_i$ of each species, and perform a summation over all the charm-hadron species in the HRGc,
\begin{equation}
\sum_{i \in \mathrm{HRGc}} q_i \tau_{i} \bar{\kappa}_{i} \dot{\nu}_q^\mu + \nu^\mu_q = \sum_{i \in \mathrm{HRGc}} q_i \kappa_{i} \cdot \nabla^\mu \alpha_q,
\label{eq:final_eom_total_diffusion_app},
\end{equation}
where we employed the definition of the normalized diffusion coefficient $\bar{\kappa}_i$, and thus $\sum_{i} q_i \bar{\kappa}_{i} = \sum_{i} q_i  \frac{\kappa_{i}}{\sum_j q_j \kappa_{j}} = 1$. Moreover, we exploited that $\nu_q^{\mu} = \nu_q^{\langle \mu \rangle}$, since by definition the diffusion current is perpendicular to the velocity. 
Eq.~\eqref{eq:final_eom_total_diffusion_app} represents the equation of motion for the charm quark, where the relaxation time and the diffusion coefficient are given by a summation of the contributions of each hadron, each weighted by the charm charge $q_i$ belonging to the charm hadron of species $i$. Plugging in the expression for charm-hadron transport coefficients reported in Eq.~\eqref{eq:taui_expression} and Eq.~\eqref{eq:kappai_expression}, one finds that the total-charm diffusion current obeys the relaxation-type equation of motion expressed by,
\begin{equation}
 \frac{D_s}{T} \sum_{i \in \mathrm{HRGc}} \frac{q^2_i  n_{i}}{\sum_j q^2_j n_{j}}  \frac{I_{i,31}}{P_{i}}   \dot{\nu}_q^\mu + \nu^\mu_q = D_s \sum_{i \in \mathrm{HRGc}} q^2_i n_{i}  \cdot \nabla^\mu \alpha_q.
\end{equation}

\section{Study of the inverse of the Reynolds number}
\label{app:reynolds_study}
\begin{figure}[h!]
    \centering
    \subfloat
    {\includegraphics[width=.45\textwidth]{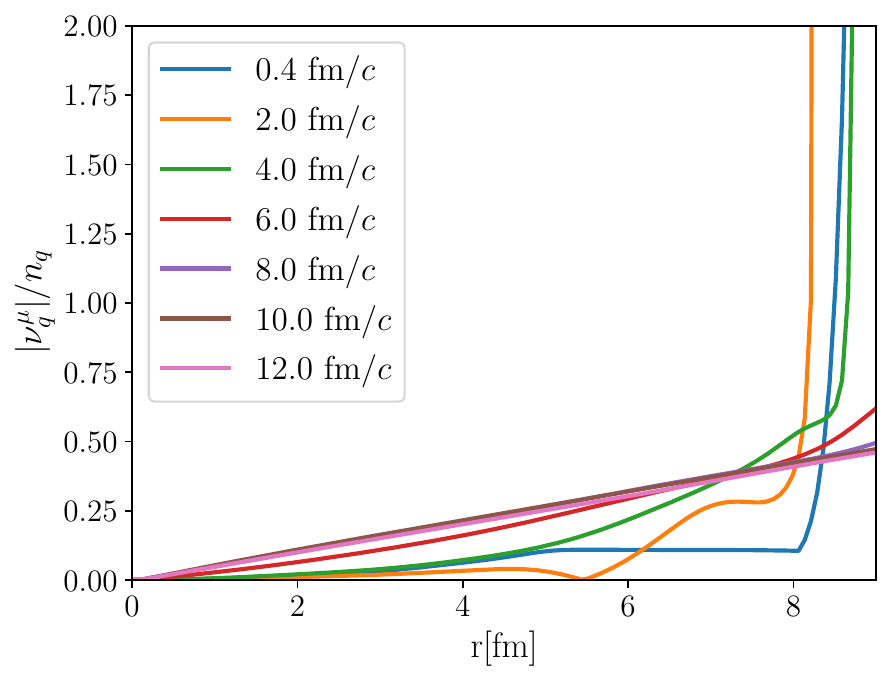}
    \label{fig:evolution_Reynolds}} \quad
    \subfloat
    {
    {\includegraphics[width=.49\textwidth]{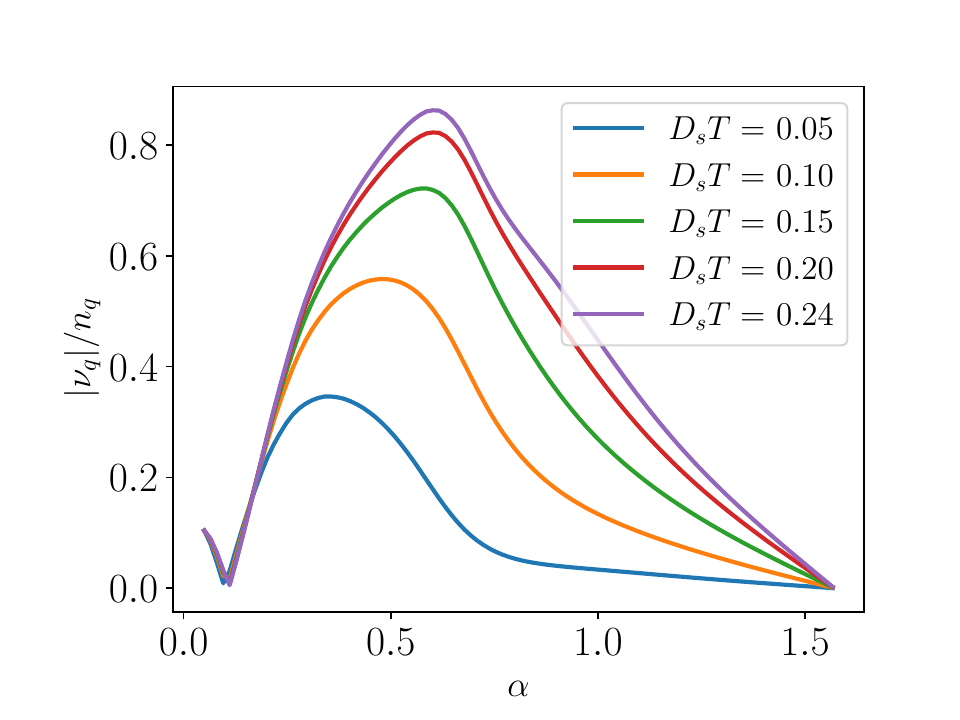}}
        \label{fig:fo_Reynolds}
    }
    \caption{Inverse of the Reynolds number during the QGP evolution at different time steps for fixed $D_sT = 0.24$ (left), and inverse of the Reynolds number at the freeze-out at different values $D_sT$.}
    \label{fig:nu_n_ratios}
\end{figure}

In order to assess the deviation from local thermal equilibrium of the charm quarks in the medium, it is possible to study the trend of the inverse of the Reynolds number as defined in Eq.~\eqref{eq:reynold}, both during the QGP phase, as well as at the freeze-out. In the left panel of Fig. \ref{fig:nu_n_ratios} we report the inverse of Reynolds number at the initial time $\tau_0 = 0.4 \ \mathrm{fm}/c$, as well as at different time steps of the evolution, for spatial diffusion coefficient $D_sT = 0.24$. The plot is shown up to a radius of $9 \ \mathrm{fm}$, corresponding approximately to the region in which the QGP fireball exists. 
As shown in the plot, after an initial growth, $\mathrm{Re^{-1}_\nu}$ decreases with time, as expected for second-order fluid-dynamic evolution, suggesting that the charm quarks approach equilibrium.

At early times and large radii, the $\mathrm{Re^{-1}_\nu}$ becomes larger than 1, indicating that in these fluid cells charm quarks are far from equilibrium. However, a portion of this region lies outside of the fireball and will therefore not contribute to transverse momentum spectra. Notice that during the fluid-dynamic regime, we evolve only moments of the distribution function. Therefore, during this stage, one cannot investigate whether the charm distribution function assumes negative values due to large out-of-equilibrium corrections.
 
In the right panel of Fig. \ref{fig:nu_n_ratios}, a study of the inverse of $\mathrm{Re^{-1}_\nu}$ on the freeze-out surface as a function of the freeze-out parameter $\alpha$ is reported. For large values of diffusion coefficient, $D_sT \sim 0.24$, the $\mathrm{Re^{-1}_\nu}$ approaches the value of 1 in the intermediate region of $\alpha$, thus possibly causing the particle distribution function to turn negative. 

\section{Study on the integrand of the distribution function}
\label{app:integrand_study}
In this section, we study the argument of the integral of the particle momentum spectrum to assess its sign and determine the regime of \pt where our model is applicable. From Eq.~\eqref{eq:Cooper:1974mv}, it is possible to analytically perform the integration over the spatial rapidity $\eta$ and the azimuthal angle $\phi$.
Considering the equilibrium contribution and neglecting the influence of resonance decays, the invariant momentum spectrum can be expressed as,
\begin{equation}
    E_\mathbf{p} \frac{d N^0_i}{d^3 \mathbf{p}} = \frac{g_i}{2 \pi^2} \int d \alpha \tau(\alpha) r(\alpha) \left[\frac{\partial r}{\partial \alpha} m_{\rm T} K_1\left({\tilde{m}_{\rm T}}u^\tau\right) I_0\left(\tilde{p}_{\rm T} u^r\right) - \frac{\partial \tau}{\partial \alpha}p_{\rm T} K_0\left(\tilde{m}_{\rm T}  u^r\right)  I_1\left(\tilde{p}_{\rm T} u^\tau\right) \right] e^{\alpha_q},
\end{equation}
where we introduced the freeze-out coordinate  $\alpha \equiv \arctan(\tau_{\rm fo}/r_{\rm fo})$, and we defined $\tilde{m}_{\rm T} \equiv \frac{m_{\rm T}}{T}$ and $\tilde{p}_{\rm T} \equiv \frac{p_{\rm T}}{T}$. The functions $I_l(x)$ and $K_l(x)$ are modified Bessel functions of order $l$ of the first and second kind, respectively. Similarly, the out-of-equilibrium contribution can be expressed as, 
\begin{equation}
\begin{split}
    E_{\mathbf p} \frac{dN^\delta_i} {d^3 \mathbf p}=& \frac{g_i}{2 \pi^2} \frac{1}{T} \frac{q_i}{\sum_j q^2_j n_j}   \int d\alpha \tau(\alpha) r(\alpha)  \times \\
    & \left\{ m_{\rm T} \frac{\partial r}{\partial \alpha} \left[p_{\rm T} \nu^r K_1(u^\tau \tilde{m}_{\rm T}) I_1(u^r \tilde{p}_{\rm T}) -m_{\rm T} \ \nu^\tau \frac{1}{2} \left( K_0(u^\tau \tilde{m}_{\rm T}) + K_2(u^\tau \tilde{m}_{\rm T})\right) I_0(u^r \tilde{p}_{\rm T}) \right]  +\right.\\
    &p_{\rm T}  \frac{\partial \tau}{\partial \alpha} \left[ m_{\rm T} \nu^\tau K_1(u^\tau \tilde{m}_{\rm T}) I_1(u^r \tilde{p}_{\rm T})
    \left.- p_{\rm T} \nu^r  \frac{1}{2}  \left( I_0(u^r \tilde{p}_{\rm T}) + I_2(u^r \tilde{p}_{\rm T})\right) K_0(u^\tau \tilde{m}_{\rm T})  \right] \right\} e^{\alpha_q}. 
    \label{eq:analytic_spectrum}
\end{split}
\end{equation}
Therefore, the integrand of the momentum spectrum reduces to functions of the transverse momentum \pt and of $\alpha$. The dependence of the fluid fields on $\alpha$ is a numerical result of the fluid dynamic evolution. 

\begin{figure}
    \centering
    \subfloat
    {\includegraphics[width=.45\textwidth]{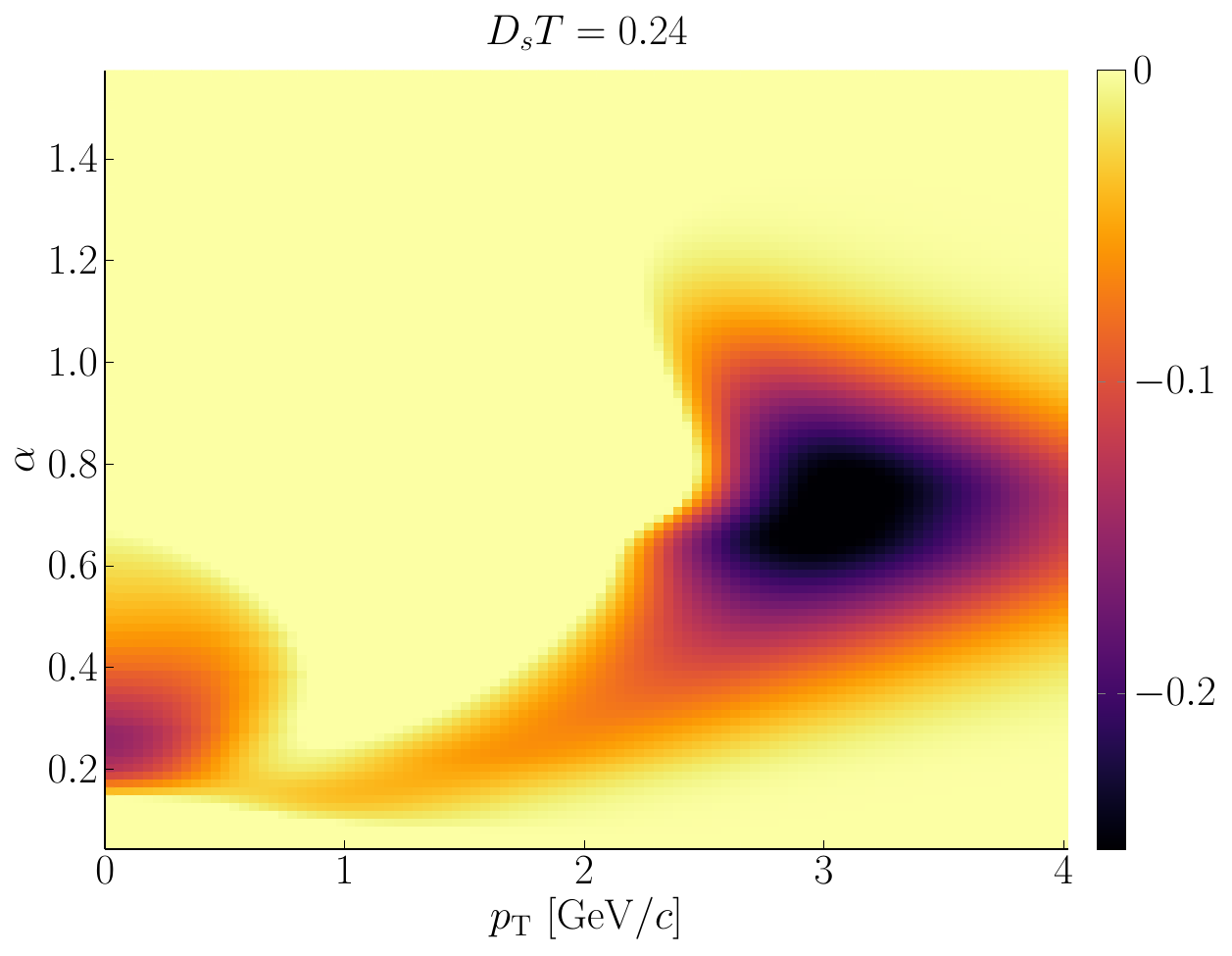}
    } \quad
    \subfloat
    {
     {\includegraphics[width=.45\textwidth]{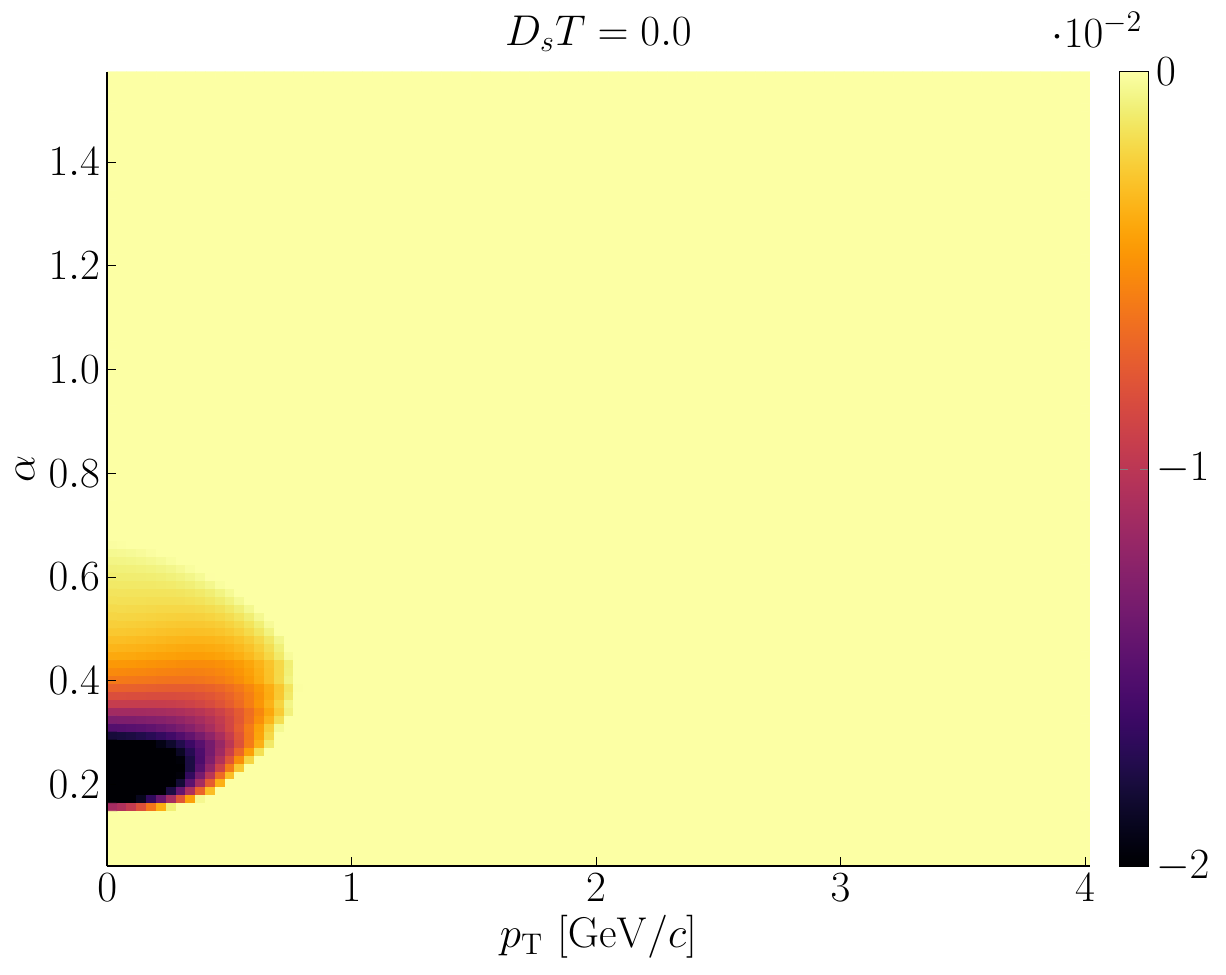}} 
     }
    \caption{Integrand of the momentum spectrum of the $\rm D^0$ meson as a function of \pt and of $\alpha$, integrated over space-rapidity and azimuthal angle, for $D_sT = 0.24$ (left panel) and for the non-diffusive case (right panel). To highlight the negative contribution, the positive contribution is reported as uniform color.}
    \label{fig:integrand_study}
\end{figure}

In Fig.~\ref{fig:integrand_study}, the integrand of the momentum spectrum of $\rm D^0$ mesons, including both the equilibrium and out-of-equilibrium contributions, is shown as a heatmap in the (\pt, $\alpha$) plane. In the left panel, the case of $D_sT = 0.24$ is considered.  We observe that, in agreement with what is reported in Fig.~\ref{fig:D0_momentum_distribution}, the integrand of the distribution function becomes predominantly negative for \pt $\gtrsim 2 \ \mathrm{GeV}/c$, causing the total momentum spectrum to be negative. However, the integrand assumes negative values at small \pt as well. Being negative within a limited region in $\alpha$, this behavior is not observed when computing the momentum spectrum, where the integration over the freeze-out coordinate $\alpha$ is performed. Remarkably, the integrand assumes negative values also at \pt $= 0$, where the contribution of the out-of-equilibrium corrections vanishes. To clarify this behavior, we report in the right panel of Fig~\ref{fig:integrand_study} the integrand of the equilibrium distribution function, corresponding to the case of $D_sT = 0$. Here, the integrand of the momentum spectrum is negative in a region at low \pt and low $\alpha$. This behavior cannot be associated with the equilibrium distribution function, since it is positive definite, but rather with the shape of the freeze-out hypersurface. As a matter of fact, since the freeze-out occurs on a constant-temperature hypersurface, the sign of ${\partial r}/{\partial \alpha}$ is not positive-definite, and specifically at low $\alpha$ it assumes negative values. We point out that a different prescription of the hypersurface, for instance, considering the freeze-out to happen at a fixed proper time $\tau_{\rm fo}$ in all the fluid cells, leads to an integrand for the equilibrium contribution that is positive in the whole (\pt, $\alpha$) plane. However, a systematic study regarding alternative parametrizations of the freeze-out hypersurface is left for future work.

\end{document}